\documentclass[article]{aa} 
\usepackage{natbib}         
\usepackage{graphicx}
\usepackage{epstopdf}
\DeclareGraphicsExtensions{.pdf,.jpeg,.png,.eps}
\usepackage{longtable}
\usepackage[switch]{lineno} 
\bibpunct{(}{)}{;}{a}{}{,} 
\usepackage{amsmath}
\usepackage{subfig}
\usepackage{xcolor}		
\usepackage{mathrsfs}

\newcommand{\ind}[1]{_{\mathrm{#1}}}

\def\Kepler{\emph{Kepler}}

\def\numax{\nu\ind{max}}\def\nmax{n\ind{max}}

\def\Dnu{\Delta\nu}

\def\Msol{M_{\odot}}

\def\largeurr{\Gamma_{0}}
\def\largeur_g{\Gamma}
\def\larg_n{\Gamma_{0} (\numax)}

\def\amp{A_{\ind{0,max}}}
\def\ampbol{A_{\ind{0,bol}}}
\def\Tbol{T_{\ind{K}}}
\def\lifetime{\tau_{0}}
\def\lifetime_tout{\tau}
\def\damp{\eta_{0}}
\def\damp_tout{\eta}
\def\velocity{\textbf{v}}
\def\puissance{P}
\def\inertie{I}

\def\Vrard{\textrm{FREQ}}
\def\Kallinger{\textrm{PROB}}

\newcommand\Teff{{T\ind{eff}}}

\newcommand\alphaM{\alpha_M}
\newcommand\alphaT{\alpha_T}

\begin{document}
\title{Amplitude and lifetime of radial modes in red giant star spectra observed by $\Kepler$}
\titlerunning{Radial mode amplitude and lifetime in red giants}
\authorrunning{M. Vrard et al.}
\author{M. Vrard\inst{1}, T. Kallinger\inst{2}, B. Mosser\inst{3}, C. Barban\inst{3}, F. Baudin\inst{4}, K. Belkacem\inst{3} and M. S. Cunha\inst{1}
} \offprints{mathieu.vrard@astro.up.pt}

\institute{Instituto de Astrof\'isica e Ci\^encias do Espa\c{c}o, Universidade do Porto, CAUP, Rua das Estrelas, 4150-762 Porto, Portugal; \email{mathieu.vrard@astro.up.pt}
\and
Institute for Astrophysics (IfA), University of Vienna, T\"urkenschanzstrasse 17, 1180 Vienna, Austria
\and
LESIA, CNRS, PSL Research University, Universit\'e Pierre et Marie Curie, Universit\'e Denis Diderot, Observatoire de Paris, 92195 Meudon cedex, France
\and
Universit\'e Paris-Sud, CNRS, Institut d’Astrophysique Spatiale, UMR 8617, 91405 Orsay Cedex, France
}

\abstract{The space-borne missions CoRoT and $\Kepler$ have provided photometric observations of unprecedented quality. The study of solar-like oscillations observed in red giant stars by these satellites allows a better understanding of the different physical processes occurring in their interiors. In particular, the study of the mode excitation and damping is a promising way to improve our understanding of stellar physics that has, so far, been performed only on a limited number of targets.}
{The recent asteroseismic characterization of the evolutionary status for a large number of red giants allows us to study the physical processes acting in the interior of red giants and how they are modify during stellar evolution. In this work, we aim to obtain information on the excitation and damping of pressure modes through the measurement of the stars' pressure mode widths and amplitudes and to analyze how they are modified with stellar evolution. The objective is to bring observational constraints on the modeling of the physical processes behind mode excitation and damping.}
{We fit the frequency spectra of red giants with well defined evolutionary status using Lorentzians functions to derive the pressure mode widths and amplitudes. To strengthen our conclusions, we used two different fitting techniques.}
{Pressure mode widths and amplitudes were determined for more than $5000$ red giants. With a stellar sample two orders of magnitude larger than previous results, we confirmed that the mode width depends on stellar evolution and varies with stellar effective temperature. In addition, we discovered that the mode width depends on stellar mass. We also confirmed observationally the influence of the stellar metallicity on the mode amplitudes, as predicted by models.}
{}

\keywords{Stars: oscillations -- Stars: interiors -- Stars:evolution -- Methods: data analysis}

\maketitle

\voffset = 1.2cm

\section{Introduction\label{introduction}}

The launches of the space missions MOST \citep{2003PASP..115.1023W}, CoRoT \citep{2006ESASP1306...33B}, and $\Kepler$ \citep{2010Sci...327..977B} have revolutionized the field of asteroseismology and especially the study of solar-like pulsators. This is particularly the case for red giant stars, MOST and CoRoT having revealed the wealth of asteroseismic observations for this kind of stars \citep[e.g.,][]{2008A&A...478..497K,2009Natur.459..398D}, then confirmed by $\Kepler$ \citep[e.g.,][]{2010ApJ...713L.176B}. Their spectra show a complicated pattern, exhibiting radial pressure modes as well as nonradial mixed modes. In cool stars with a convective envelope, pressure modes are the signature of acoustic waves excited by turbulent convection in the outer stellar layers. They carry information on the internal structure of the envelope and on the different physical processes occurring inside these objects. Their spectra follow the so-called universal red giant oscillation pattern \citep{2011A&A...525L...9M}, which describes the regularity of the pressure mode pattern characterized by two quantities: the frequency  of maximum oscillation power ($\numax$) and the mean frequency difference ($\Dnu$) between consecutive pressure modes of the same angular degree. In combination with effective temperature ($\Teff$), these global seismic parameters are used to precisely constrain the stellar mass and radius through scaling relations \citep[e.g., ][]{2010A&A...522A...1K}.

Mixed modes, noticed in red giants by \citet{2010ApJ...713L.176B}, then identified by \citet{2011Sci...332..205B}, are the signature of waves behaving as acoustic waves in the stellar envelope and as gravity waves in the core. Therefore, they carry unique information on the red giant core properties, which allows one to distinguish between core-helium burning red giants (clump stars) and hydrogen-shell burning red giants \citep[RGB stars; ][]{2011Natur.471..608B,2011A&A...532A..86M,2012A&A...540A.143M}. The identification of the stars evolutionary status was then performed on a large number of objects \citep{2012A&A...541A..51K,2013ApJ...765L..41S,2017MNRAS.466.3344E,2017MNRAS.469.4578H}. Recent advances \citep{2014A&A...572L...5M,2017A&A...600A...1M,2016A&A...588A..87V} make it even possible to precisely follow the evolution of the stellar core mass during the red giant phase.
\newline

Among the physical processes occurring inside red giants, those responsible for mode excitation and damping are most difficult to characterize since they involve non-adiabatic physical processes. Constraints on the excitation and damping mechanisms can be obtained by measuring the mode amplitudes and lifetimes, the latter being directly related to the mode widths. The physical mechanisms behind mode damping are still poorly identified. Several studies were conducted to characterize the dominant one in solar-like pulsators, leading to different conclusions \citep[e.g.][]{1991ApJ...374..366G,1992MNRAS.255..639B,2006ESASP.624E..97D}. The absence of a global agreement on this subject makes theoretical predictions difficult. A first attempt to characterize the scaling relations between pressure mode damping and stellar parameters, such as the effective temperature, was done by \citet{2009A&A...500L..21C}, but these theoretical predictions were later challenged by observations \citep{2011A&A...529A..84B,2012ASPC..462....7H}. However, recent theoretical studies taking the latest observational results into account seem to point towards the perturbation of turbulent pressure being the main physical mechanism behind mode damping in solar-like pulsators \citep{2012A&A...540L...7B,2017MNRAS.464L.124H}. Several other studies intended to constrain the mode damping based on the observed mode lifetimes, but they are mainly focused on main-sequence and subgiants stars \citep{2012A&A...537A.134A,2014A&A...566A..20A,2017ApJ...835..172L}. Few studies have been conducted for red giants and they were restricted to a limited number of targets \citep{2011A&A...529A..84B,2012ApJ...757..190C,2015A&A...579A..83C,2017MNRAS.472..979H}. A thorough measurement of the pressure mode lifetime for a large number of red giants is still to be performed.
The physical mechanism responsible for pressure mode
excitation, better understood than the damping, has been identified as
the Reynolds stresses, which are induced by turbulent
convection \citep[e.g.][]{1992MNRAS.255..639B,2006A&A...460..183B}.

Several studies propose scaling relations between mode amplitudes and stellar parameters \citep[e.g. ][]{1995A&A...293...87K,1999A&A...351..582H,2007A&A...463..297S,2011A&A...529L...8K}. 
A scaling relation with mode amplitude varying as $(L/M)^{0.7}$ was predicted for main-sequence stars, where $M$ is the mass of the star and $L$ its luminosity. \citet{2011A&A...529L...8K} introduced afterward the fact that the relative variation of the luminosity not only scales with $L/M$, but also with the mean lifetime of the modes and with the large separation. 
It was later extended to red giant stars by \citet{2012A&A...543A.120S}.
This relation has been tested on red giants by numerous studies \citep{2010A&A...517A..22M,2010ApJ...723.1607H,2011A&A...529A..84B,2012A&A...537A..30M,2013MNRAS.430.2313C,2014A&A...570A..41K}, which found scalings varying typically as $L^{0.8}/M^{1.2}$. This difference compared to a scaling as a power law of $L/M$ indicates the relevance of examining the role of an additional term such as mode lifetimes in the mode amplitude.
\newline

Here, we analyze the pressure mode amplitudes and lifetimes in a large number of red giant oscillation spectra. Using several thousands oscillation spectra from the main \emph{Kepler} mission, we aim to investigate the dependence between the mode parameters and the global stellar parameters and follow their variation with the stellar evolution. We then compare our results with previous measurements and theoretical predictions. The layout of the paper is as follows. In Section $2$, we recall how the observed mode width and amplitude are linked with the mode damping and excitation. In Section $3$, we describe the two fitting methods we used. The analysis of the measured mode widths and amplitudes is conducted in Section $4$ and $5$, respectively. Section $6$ is devoted to conclusions.

\section{Observational constraints on mode damping and excitation}\label{Introduction}

\subsection{Mode damping}

Mode widths $\largeur_g$, defined as the mode FWHM, are directly related to mode lifetime $\lifetime_tout$ and, consequently, to the considered mode damping rate $\damp_tout$ following the relation \citep[e.g.,][]{2015EAS....73..111S}

\begin{equation}
   \largeur_g = \frac{1}{\pi\lifetime_tout} = \frac{\damp_tout}{\pi} .
   \label{Equation:width-damping}
\end{equation}
Measurements of the mode width therefore directly constrain the mode damping.

\subsection{Mode amplitude}

The radial mode amplitude is related to the energy injected into the modes and can therefore be linked to the local mean squared velocity variation ($\velocity$) of the oscillations \citep[see for example Eq. (4.15) of][]{2015EAS....73..111S}, which is given by \citep[e.g. ][]{2009A&A...506...57D,2013EPJWC..4303008S}

\begin{equation}
   \velocity^{2} = \frac{\puissance}{2 \damp_tout M \inertie} ,
   \label{Equation:velocity-power}
\end{equation}
where $\puissance$ is the time-averaged power stochastically supplied to the mode and $\inertie$ is the mode inertia. The mode amplitude is then linked to the $\puissance$ parameter and the mode damping.

\section{Fitting procedure\label{Principle}}

Our aim is to assess the dependence of the pressure mode widths and amplitudes on the stellar fundamental parameters. To achieve this, we use two different automated methods. The first is based on a frequentist fitting and the second follows a Bayesian approach. The independent development of these methods allows us to strengthen the conclusions regarding the different discovered dependencies.

\subsection{Data}	\label{sec:data}

Long-cadence data from $\Kepler$ up to the quarter Q$17$ are used, which correspond to $44$ months of photometric observations. The red giant sample was selected based on the APOKASC Catalog \citep{2014ApJS..215...19P}, which resulted from an asteroseismic and spectroscopic joint survey of $\Kepler$ targets. This survey includes about $6300$ red giants. In the following, we focus our analysis on a subset of $5523$ stars with a signal-to-noise ratio sufficient to determine their evolutionary states using the following techniques.

The evolutionary status is determined for each star using the method of \citet{2012A&A...541A..51K}, which is based on the evaluation of the difference between clump and RGB stars in their pressure mode pattern. This method is completed, when possible, by the one of \citet{2016A&A...588A..87V}, which uses the mixed-mode pattern to distinguish clump and RGB stars.
\newline

As stated above, the spectra of red giant stars exhibit pressure modes as well as mixed modes. Because of the nature of gravity waves, radial modes ($\ell=0$) are not affected by the coupling between gravity and pressure waves, contrary to higher angular degree modes, and they are not split by rotation. We therefore focus our study on radial modes since they are the only pure pressure modes present in the red giant oscillation spectra.

\subsection{General description}

The two mode-fitting methods show important similarities in their execution, with four main steps.

\begin{itemize}
  \item Determination of the seismic global parameters ($\Dnu$ and $\numax$) as well as the characterization of the granulation background.
  \item Refinement of the estimate of the large separation through the use of the pressure mode frequency positions. These modes indeed follow a precisely described pattern that has been characterized in previous works \citep[e.g.,][]{2010A&A...509A..77K,2011A&A...525L...9M}. The use of this pattern allows a precise adjustment of $\Dnu$ and to obtain an approximation of the position of the radial modes.
  \item Identification of the radial modes in the spectra with an estimation of their frequency positions.
  \item Fitting of the radial and nearby modes on top of the previously estimated background by Lorentzian profiles, (one per mode) which are defined by
  \begin{equation} \label{eq:lor}
\mathcal{L}(\nu) = \sum_k \frac{A_k^2/(\pi\Gamma_k)}{1+\left(\frac{2(\nu-\nu_k)}{\Gamma_k}\right)^2},
  \end{equation}
  where $A_{k}$ is the amplitude of the $k^{th}$ mode, $\nu_{k}$ is its oscillation mode frequency, and $\Gamma_{k}$ is its mode linewidth (FWHM, see Eq. \ref{Equation:width-damping}). The radial mode widths, heights, amplitudes and frequencies are then extracted for all significant modes.
\end{itemize}

\subsection{First method: frequentist approach ($\Vrard$)}

The method $\Vrard$ realizes the above mentioned steps as follows.
\newline

The first step is performed by the use of the envelope autocorrelation function as described in \citet{2009A&A...508..877M}. It allows, for a given spectrum, the determination of a first estimate of $\Dnu$ and $\numax$ as well as the background component parameterized as a power law \citep[Eq. 3 of][]{2012A&A...537A..30M}. 

The second step is carried out by means of the universal pattern as described in \citet{2011A&A...525L...9M}, which refines the value of $\Dnu$ from the autocorrelation function and allows us to estimate the frequency position of the radial modes.

The third step is realized by smoothing the power density spectrum with a low-pass filter with a width equal to $\Dnu/100$ and locating the local maximas within one tenth of $\Dnu$ around the estimated radial mode positions. These local peaks are considered to be significant if their heights exceed a threshold corresponding to the rejection of the pure noise hypothesis with a confidence level of $99.9\%$. Among those peaks, the $\ell = 0$ and $\ell = 2$ doublets can easily be identified as the modes with the highest mode heights and largest mode widths. If there are more than two peaks exhibiting a height and width value higher than half the height and half the width of the highest and widest peak, respectively, then the radial mode is identified as the one closest to the frequency corresponding to the first estimation of the radial mode frequency position.

\begin{figure}[t]              
  \includegraphics[width=9cm]{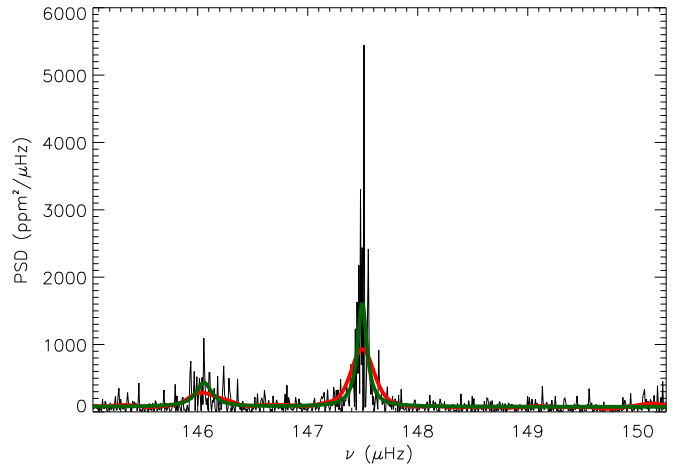}
  \caption{Power density spectrum for the star KIC3758505 (black line) and the range of frequencies corresponding to one tenth of $\Dnu$. The red line represent the smoothed spectrum. Our best fit (green line) consists of two Lorentzians and a power law background component with the peaks at about $147.5$ and $146$ $\mu$Hz being a radial and $\ell = 2$ mode, respectively.}
  \label{fig:ajustement}
\end{figure}

The last step is performed by fitting the radial and the nearby modes with Lorentzian profiles (Fig. \ref{fig:ajustement}) using the maximum likelihood estimator technique described in \citet{1994A&A...289..649T} and \citet{2014aste.book..123A}. The fit consists of a background component parameterized as a power law \citep[Eq. 3 of][]{2012A&A...537A..30M} and several Lorentzian profiles, one for each modes (see Eq. \ref{eq:lor}), following the model of \citet{2010AN....331.1016B} and \citet{2015A&A...579A..84V}. This allows us to extract the frequency, height, width, and amplitude of all significant modes, including radial and quadrupolar modes. The use of a power law as a model for the background has already been proven to be acceptable for small portions of the spectrum \citep{2012A&A...537A..30M}.
\newline

We use this method on the $\Kepler$ data previously described but select only the stars for which we can extract at least three radial modes with uncertainties on their measured parameters lower than $10 \%$. This criterion is satisfied by $5221$ stars.

Since the method $\Vrard$ is fully automated, spurious measurements are possible. It sometimes happens that a radial mode has a lower amplitude than its nearby $\ell=2$ mode. If so, confusion in the mode identification is possible and the $\ell=2$ mode is identified as a radial mode. Such incidences are, however, quite rare since the quadrupole modes typically have an amplitude two times lower than the radial modes \citep{2012A&A...537A..30M}. Following on the different tests we performed, we confirm that this kind of confusion happens in less than $1\%$ of all investigated stars.

Several previous studies also pointed out that there is a significant impact from the background parameterization on the mode fitting and therefore on the mode widths and amplitudes \citep[e.g. ][]{2014A&A...566A..20A}. However, it was previously demonstrated that the use of a power law as a background model offer enough precision for estimating the background in the limited frequency range where oscillations are observed \citep{2012A&A...537A..30M}. Nevertheless, we can not rule out the existence of global systematic effects connected to the selection of this particular background model.
\newline

The complete data sets listing the mode-fitting results that were obtained with this method for each star are available at the CDS (see Appendix \ref{Section: CDS_data_individualstars}).

\subsection{Second method: probabilistic approach ($\Kallinger$)}

For the second approach, the first step is done by using the approach of \cite{2014A&A...570A..41K} that characterizes the granulation background in the vicinity of the oscillation power excess and models the global shape of the power density spectrum (PDS) with the superposition of two super-Lorentzian functions\footnote{Super-Lorentzian functions are similar to Lorentzians but with a different exponent in their denominator \citep[see e.g.,][]{2014A&A...570A..41K}.} and a Gaussian centered on $\numax$, where the latter serves as a proxy for the oscillation power.

The second and third steps are performed by means of a fit based on the following relation, which mimics the second-order asymptotic expansion
  \begin{equation} \label{eq:UP_Kal}
\nu_{n,\ell} = \nu_c+\Dnu\left[n-n_c+\frac{\alpha}{2}(n-n_c)^2\right]-\delta\nu_{0\ell},
  \end{equation}
  where $\nu_c$ is the frequency of the radial mode closest to $\numax$, $n_c$ its radial order, $\delta\nu_{0\ell}$ the small separation and $\alpha$ a curvature term. This relation provides relevant estimates for $\nu_c$, $\Dnu$, and the other parameters and allow a precise estimation of the position of the pure pressure $\ell=0$ and $2$ modes in five radial orders around $\numax$. Tests have shown that we can re-construct the full observable $\ell=0$ and 2 mode pattern of red giants in all evolutionary stages covered by the $\Kepler$ observations to better than about 2\% of $\Dnu$.

The last step consists in fitting the individual radial modes with Lorentzian profiles and rating their statistical significance. To do so, we fit the sum of two Lorentzian profiles, following Eq. (\ref{eq:lor}), on top of the previously defined background to the PDS in the frequency range of $-$1.5 to +0.5 times $\delta\nu_{02}$ around the estimated position of a given radial mode. For the fit we use the Bayesian nested sampling algorithm \textsc{MultiNest} \citep{2009MNRAS.398.1601F}, which delivers posterior probability distributions for the frequency ($\nu_k$), rms amplitude ($A_k$), and width ($\Gamma_k$) of the $k$th profile, from which we determine the best-fit parameters and their $1\sigma$-uncertainty. A problem in this approach is the potential presence of mixed $\ell=1$ modes in the close vicinity of the radial or quadrupole mode, which would distort the fit. This is tested by dividing the PDS by the initial fit, which then represents the residuals  in terms of a height-to-background ratio (HBR). If we find sharp peaks exceeding a HBR of eight, they are eliminated from the PDS and the fit is redone.
\newline

To rate the significance of the fitted modes we compare the above fit to fits including only one Lorentzian profile (i.e., only the $\ell=0$ or $\ell=2$ mode is detectable in the given frequency range) and no profile at all (i.e., only an average value). A scenario is then considered to be statistical significant if its model evidence (as delivered by \textsc{MultiNest}) is at least ten times\footnote{In probability theory an odds ratio of 10:1 is considered as strong evidence \citep{jeffreys98}.} larger than the combined model evidence of the other scenarios. We repeat this procedure for $\ell=0$ and 2 doublets for relative radial orders of up to $\pm$5, which gives in the best case a sequence of 11 radial modes, for which we can determine their frequencies within typically 5-10\,nHz in the center of the power excess and some 10\,nHz at the wings. The median relative uncertainty of the mode width in the central region of the power excess is about 12\%.
\newline

The method $\Kallinger$ represents a reliable approach to determine the radial mode parameters of red giants in a fully automatic way. Extensive tests have shown that the misidentification of a mode is very rare. It is, however, quite expensive in terms of computation time (about 15\,min for a single star on a personal computer). The complete data sets listing the mode-fitting results from this method will be published in a forthcoming paper (Kallinger et al. in preparation).

\subsection{Computation of the global radial mode width and the bolometric amplitude}

For each star and each method, the global radial mode width $\larg_n$ is computed as the weighted mean of the three pressure modes closest to $\numax$, where we use the mode amplitude as weight. Such a definition minimizes systematic effects due to the possible spurious measurements previously mentioned. Moreover, at these frequencies, the observed mode widths are expected to exhibit almost identical values \citep{1999A&A...351..582H,2014A&A...566A..20A}. $\larg_n$ then corresponds to the nearly uniform mode widths, which are observed close to $\numax$.
\newline


Concerning the mode amplitude, we first determine the average maximum mode amplitude $\amp$ for a given star as the mean amplitude of the three radial modes closest to $\numax$, following \citet{2011A&A...529A..84B}. Before analyzing the amplitude variations, we bolometrically correct the $\Kepler$ observations according to
\begin{equation}
   \ampbol = \amp \left(\frac{\Teff}{\Tbol}\right)^{0.80} ,
   \label{Equation:velocity-power}
\end{equation}
with $\Tbol = 5934 K$ \citep{2011A&A...531A.124B}. This correction accounts for the wavelength dependence of the photometric variation integrated over the $\Kepler$ bandpass. The complete data are available at the CDS (see Table \ref{tab:CDS_file}).
\newline

\begin{table*}[t]
\caption{Global seismic parameters \label{tab:CDS_file}}

\begin{center}
\begin{tabular}{ccccccccccccc}
 \hline
KIC number & $\Dnu$ ($\mu$Hz) & $\numax$ ($\mu$Hz) & $\larg_n$ ($\mu$Hz) & $\delta\larg_n$ ($\mu$Hz) & $\ampbol$ (ppm) & $M/M_\odot$   & $\delta M/M_\odot$ & evolution\\
\hline
1027337  &  6.95   &  75.1   &   0.099   &   0.0111   &   60.53   &   1.49   &    0.044    &   1\\
1160789  &  3.61   &  25.3   &   0.156   &   0.0301   &   145.66   &   0.79   &    0.023    &   2\\
1161447  &  4.22   &   36.8   &   0.158   &   0.0274   &   74.28   &   1.34   &    0.039    &   2\\
1161618  &  4.11   &   33.9   &   0.139   &   0.0239   &   111.61   &   1.14   &    0.033    &   2\\
\hline
\end{tabular}

\end{center}

Seismic parameters, as described in the text, of four stars among the $5523$ that were considered in this study. The full table for the whole data set is available at CDS. The columns correspond to, from left to right, the star KIC number, its large separation ($\Dnu$), its frequency of maximum oscillation power ($\numax$), its global radial mode width ($\larg_n$), the $1\sigma$ uncertainties on this parameter, its bolometric mode amplitude $\ampbol$, its stellar mass ($M$) and the $1\sigma$ uncertainties on this parameter. The 'evolution' value on the last column provides the evolutionary stage: $1$ for RGB stars and $2$ for clump stars.

\end{table*}

\subsection{Comparison of the methods}\label{Methods_bias}

\begin{figure}[t]              
  \includegraphics[width=9cm]{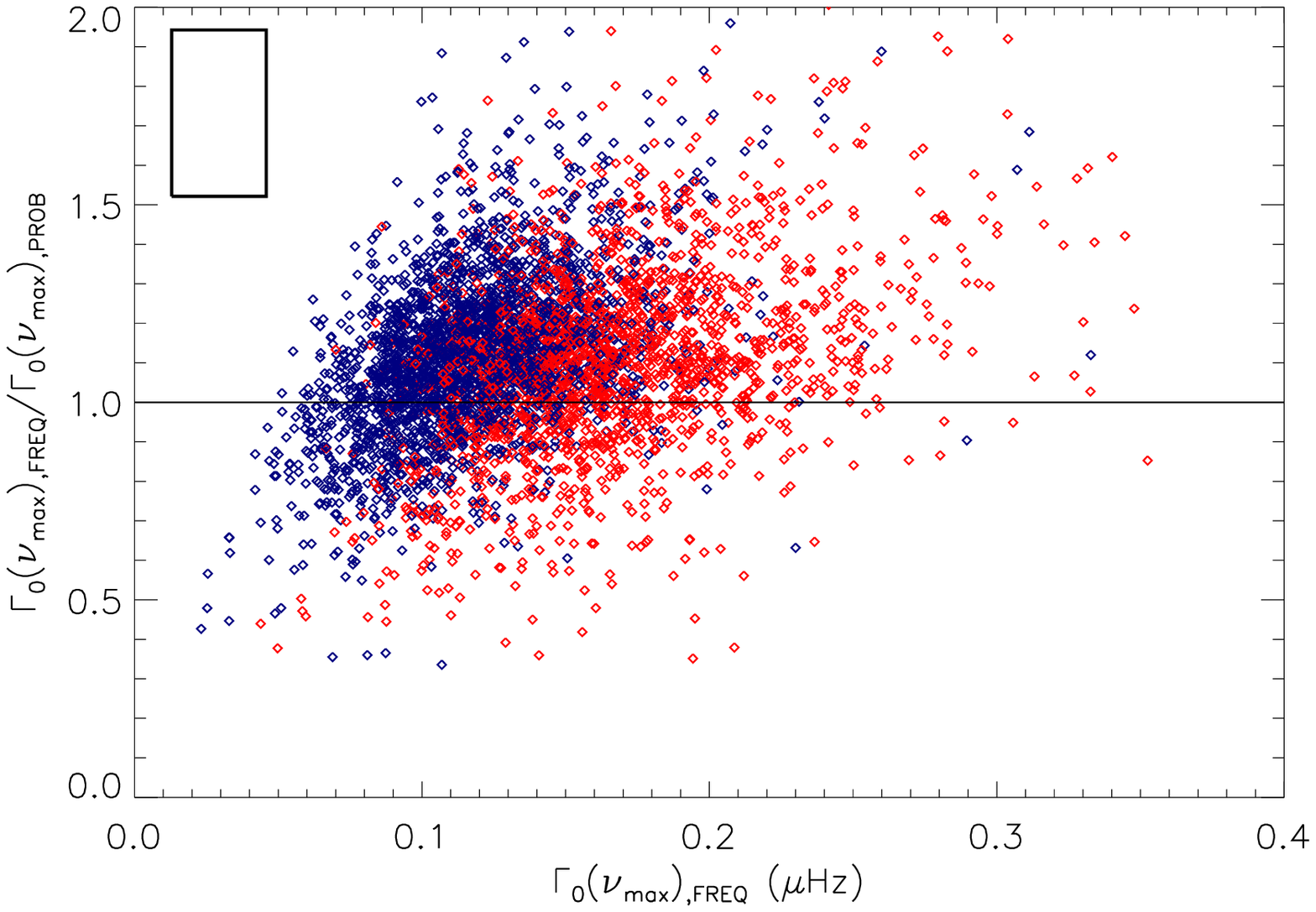}
  \includegraphics[width=9cm]{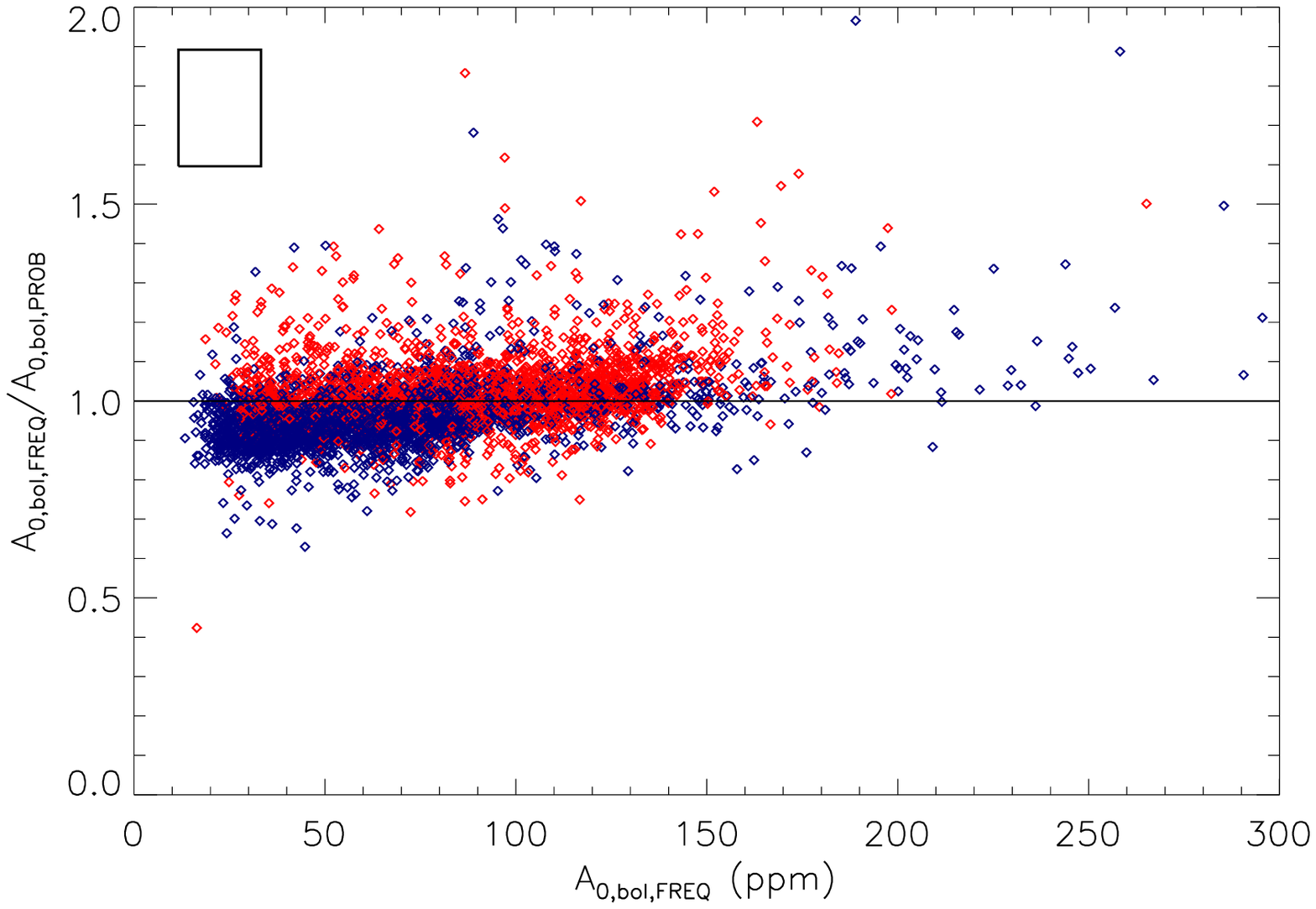}
  \caption{\textit{Top:} ratio between the global radial mode widths $\larg_n$ obtained with the $\Vrard$ method and those obtained with the $\Kallinger$ method as a function of the global radial mode width $\larg_n$ obtained with the $\Vrard$ method.
\textit{Bottom:} ratio between the bolometric amplitudes $\ampbol$ obtained with the $\Vrard$ method and those obtained with the $\Kallinger$ method as a function of the bolometric amplitudes $\ampbol$ obtained with the $\Vrard$ method. 
For both panels, the black line represents the 1:1 relation and the mean uncertainties are given by the black rectangle in the upper left corner. The red diamond and blue triangle symbols indicate clump and RGB stars, respectively.}.
  \label{fig:Compar_width}
\end{figure}

In order to evaluate the consistency of our results, we compare the mode parameters for the two methods. Figure \ref{fig:Compar_width} gives a comparison between the radial mode widths and amplitudes, respectively, determined with the different methods.

The results are consistent with each other for both parameters and their observed dispersion is consistent with the error bars. However, we can see that the $\Vrard$ $\larg_n$ values are globally overestimated by about $10\%$ with respect to the $\Kallinger$ values. This deviation is independent of the evolutionary stage as well as of the mass of the stars.
\newline

In order to more precisely characterize this bias, we perform a linear fit on the different values we found for the two methods. The results are the following:

\begin{equation}
   \left\{
       \begin{array}{l}
           \larg_n{}\ind{,\Vrard} = (0.91\pm0.02)\larg_n{}\ind{,\Kallinger} \\
           +  0.023\pm0.002    \\
           \\
           \ampbol{}\ind{,\Vrard} = (1.04\pm0.02)\ampbol{}\ind{,\Kallinger} - 3.8\pm0.35
       \end{array}
   \right.
   \label{Equation:fit_biais}
\end{equation}

The slopes we find for the fits are close to one for both parameters, indicating that the results of the two methods scale similarly with the parameters under consideration. The small slope for $\larg_n$ could indicate the existence of a small trend but, the measured values being concentrated in a small range of $\larg_n$, does not allow to draw firm conclusions. Nevertheless, in the case of mode widths, the constant in the fit corresponds to about $10\%$ of the line width values, reflecting the bias towards larger values of the line widths obtained with the $\Vrard$ method.

In order to estimate the impact of the mode width bias, we compute the mean quadratic deviations between the $\larg_n$ of the two methods. We find that the value of this parameter corresponds to $0.036$ $\mu$Hz, which is smaller than the mean uncertainty on the mode width ($0.042$ $\mu$Hz). We therefore conclude that this bias does not significantly impact the further interpretation of the results.

\section{Mode widths}\label{sec:mode_width}

In this section, we show the results we obtain for the radial mode widths, compare them to theoretical predictions and previous results on red giants, and discuss the implications.

\subsection{$\larg_n$ variation with the stellar evolution: evidence for a mass dependence}

Since the large separation is directly related to the mean stellar density, we can use $\Dnu$ to estimate the variation of $\larg_n$ with stellar evolution along the RGB. Moreover, we can search for differences between clump and RGB stars at same $\Dnu$ (i.e., same mean stellar density). To estimate the influence of the stellar mass on $\larg_n$, we determine the mass from scaling relations for the global oscillation parameters ($\Dnu$ and $\numax$) and the effective temperature \citep[taken from the APOKASC measurements, see][]{2014ApJS..215...19P}.
\newline

\begin{figure}[t]              
  \includegraphics[width=9cm]{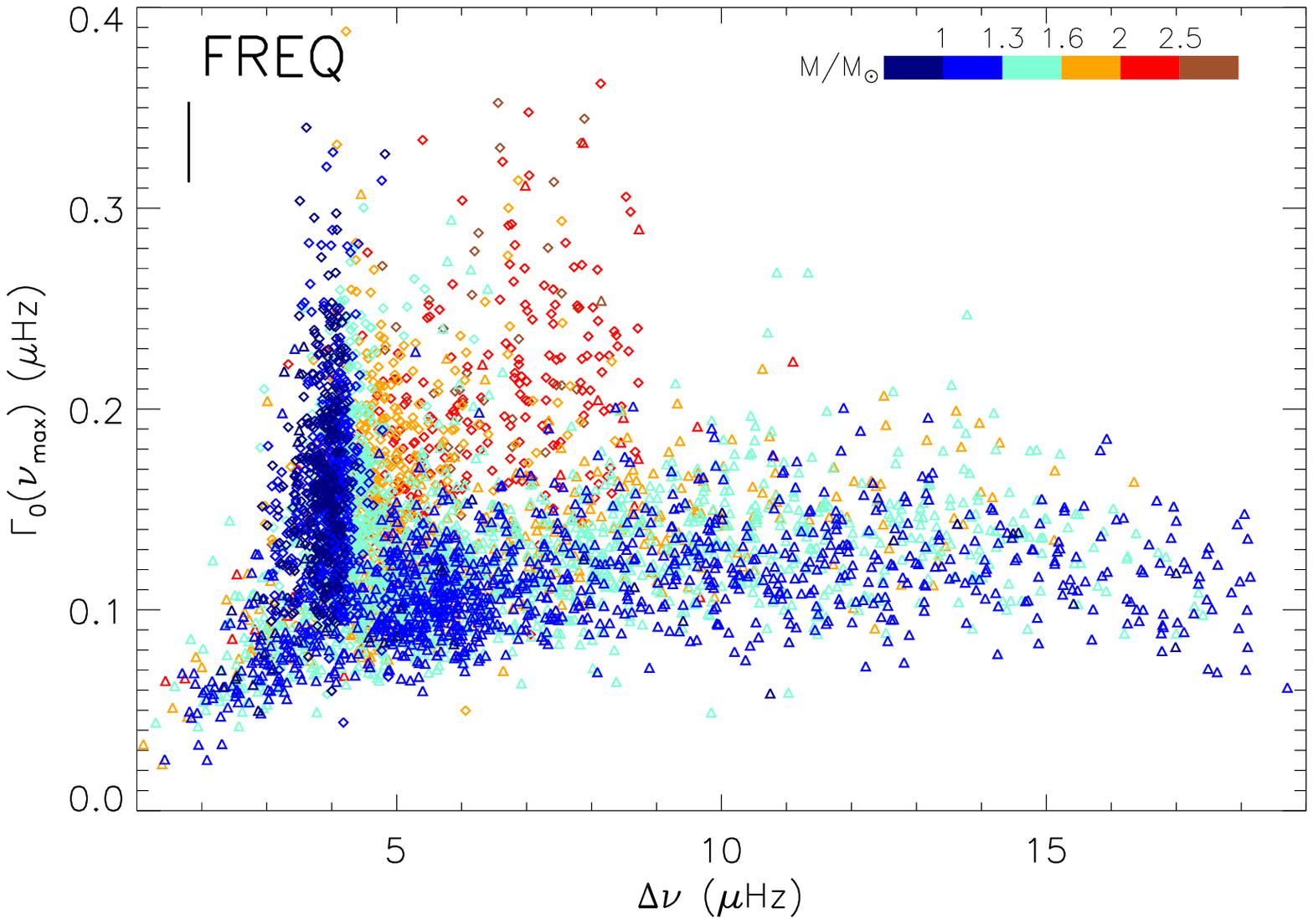}
  \includegraphics[width=9cm]{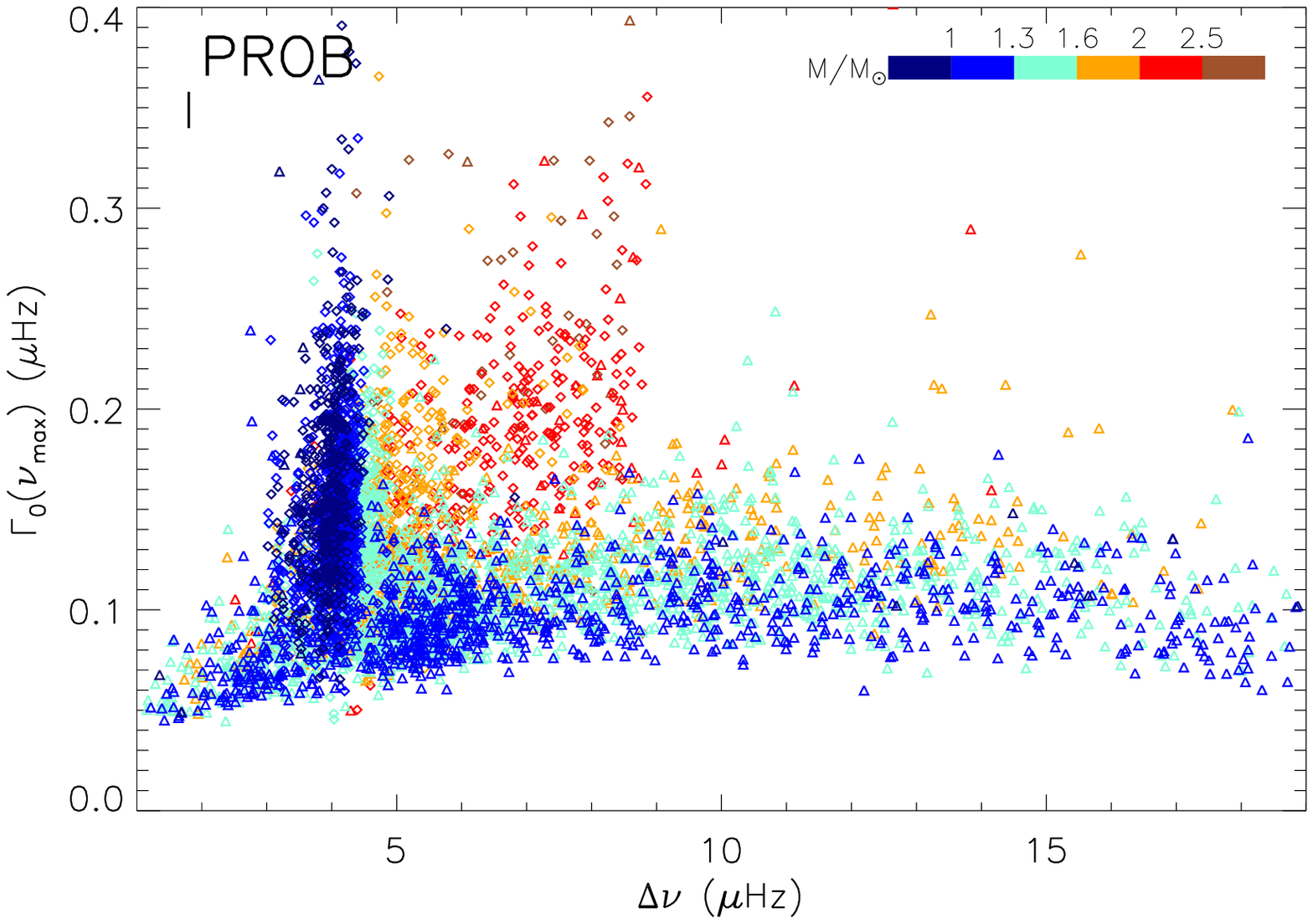}
  \caption{Global radial mode width ($\larg_n$) as a function of the large separation ($\Dnu$) for the $\Vrard$ method (top) and for the $\Kallinger$ method (bottom), with the stellar mass color-coded. The diamond and triangle symbols indicate clump and RGB stars, respectively. The mean uncertainty in $\larg_n$ is given by the vertical black line in the upper left corner.}
  \label{fig:largeur_mass}
\end{figure}

The dependence of $\larg_n$ on the large separation is shown in Fig. \ref{fig:largeur_mass}, where we find very similar results for both methods. The main difference is a larger dispersion for the $\Vrard$ method, which is due to larger uncertainties. We also find a small but clear mass dependence, regardless the stars evolutionary status. This is particularly obvious for helium burning stars. The values of $\larg_n$ for clump stars show indeed a clear dependence on $\Dnu$, especially in the secondary clump, which is a consequence of the mass dependence. At fixed $\Dnu$, a star with a higher mass has a higher $\larg_n$ value. The influence of the stellar mass on the pressure mode widths was not widely considered by previous theoretical studies. It is then difficult to assess the implications of such a behavior. This result suggests that more theoretical work is needed to understand the contributions of the different physical mechanisms behind mode damping.

We also note that, for stars with similar masses and the same large separation, the global radial mode widths in the clump are significantly larger than those on the RGB as was previously mentioned by \citet{2012ApJ...757..190C}. Given the clear difference in behavior between the two evolutionary states, we will separate them in the analysis that follows.
\newline

Finally, two minor features can be observed. Firstly, low $\larg_n$ values are observed for evolved RGB stars (bottom part of Fig. \ref{fig:largeur_mass}), which can be explained by their low $\Teff$ (see Sec.\,\ref{Temperature_dependence}). 
The second one concerns the way $\larg_n$ evolves along the RGB: both methods show a non-monotonous variation of this parameter with $\Dnu$. It certainly indicates that the large separation is not a good proxy of the stellar evolution for studying the mode width. 
Further studies are needed in order to understand this particular behavior.

\subsection{The $\larg_n$-$\Teff$ relation\label{Temperature_dependence}}

According to theoretical predictions, the stellar linewidths should strongly depend on the stellar effective temperature \citep[e.g., ][]{2009A&A...500L..21C,2012A&A...540L...7B}. Such a strong correlation has already been confirmed for main-sequence and subgiant solar-like pulsators \citep{2011A&A...529A..84B,2012A&A...537A.134A}. For red giants, however, the radial mode widths seem to be less strongly correlated with $\Teff$ \citep{2011A&A...529A..84B,2012ApJ...757..190C,2015A&A...579A..83C} or even independent of it \citep{2017MNRAS.472..979H}. As an example, to take mode widths measurements from both main-sequence and red giant stars into account, \citet{2012ApJ...757..190C} had to introduce an exponential dependence of the mode widths with $\Teff$. Here, we analyze the $\larg_n$-$\Teff$ correlation in order to confirm or infirm the previous results. For this, we used the effective temperatures listed in the latest APOKASC Catalog \citep[see e.g., ][]{2014ApJS..215...19P}.
\newline

\begin{figure*}[t]              
  \includegraphics[width=9cm]{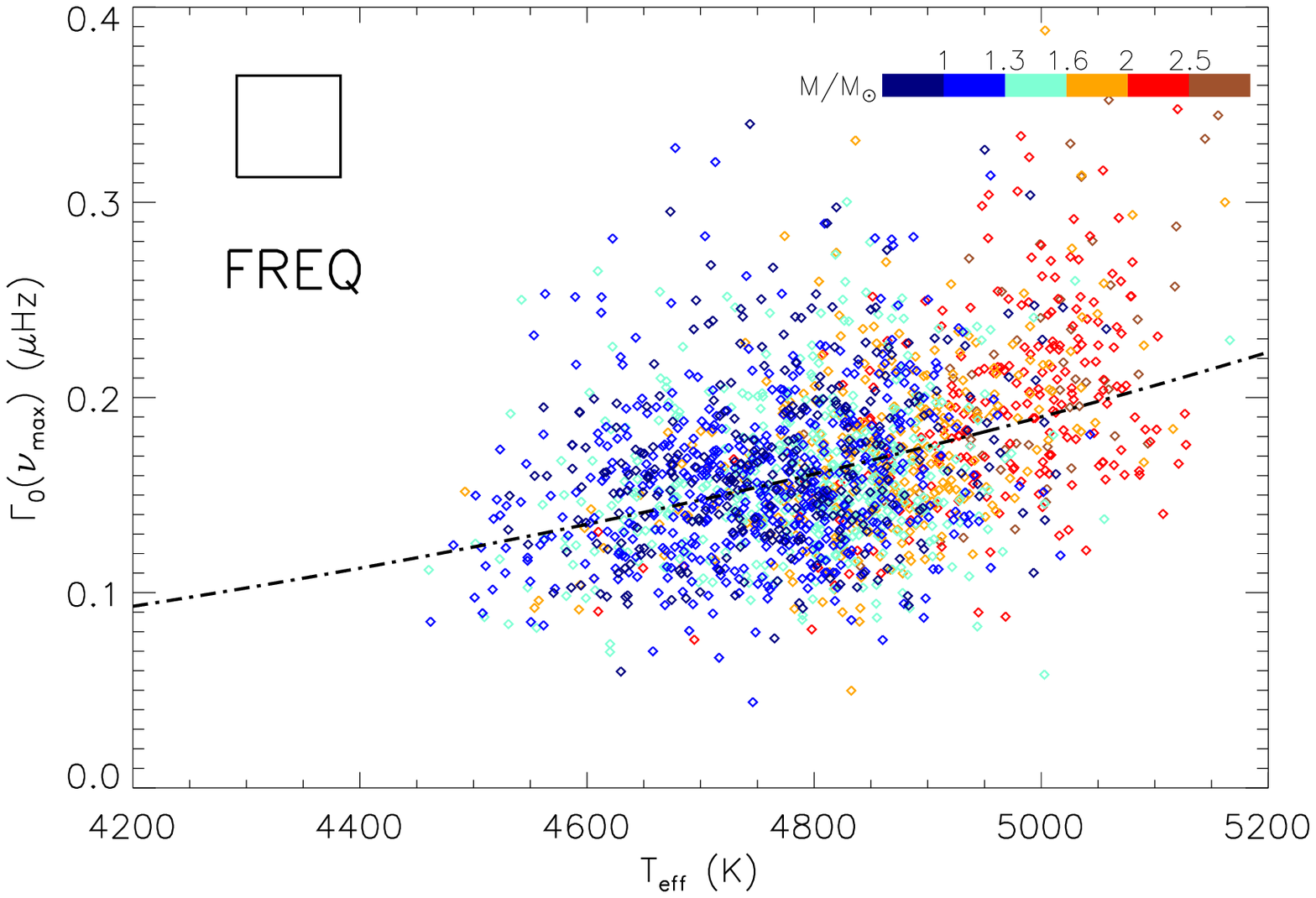}
  \includegraphics[width=9cm]{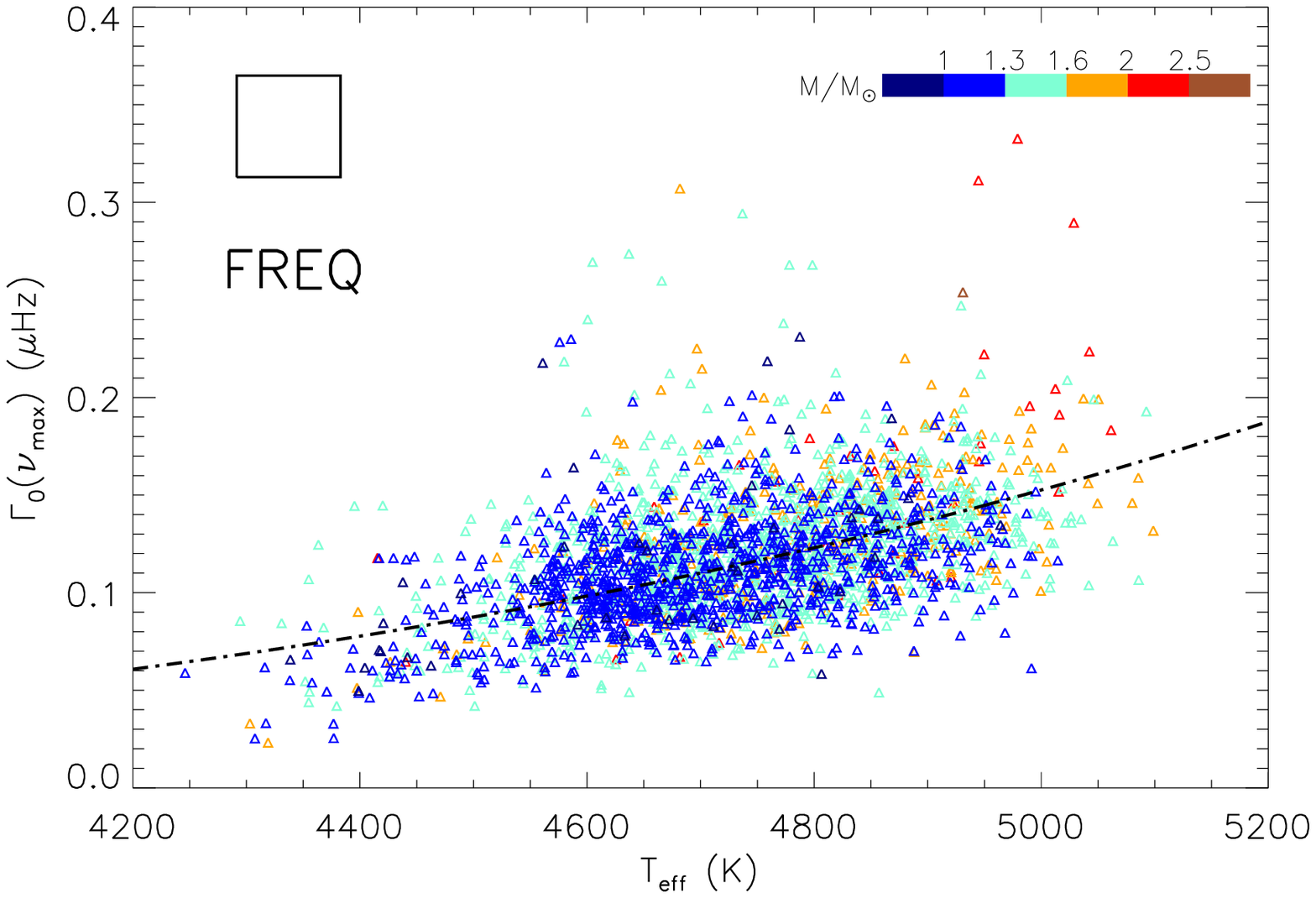}
  \includegraphics[width=9cm]{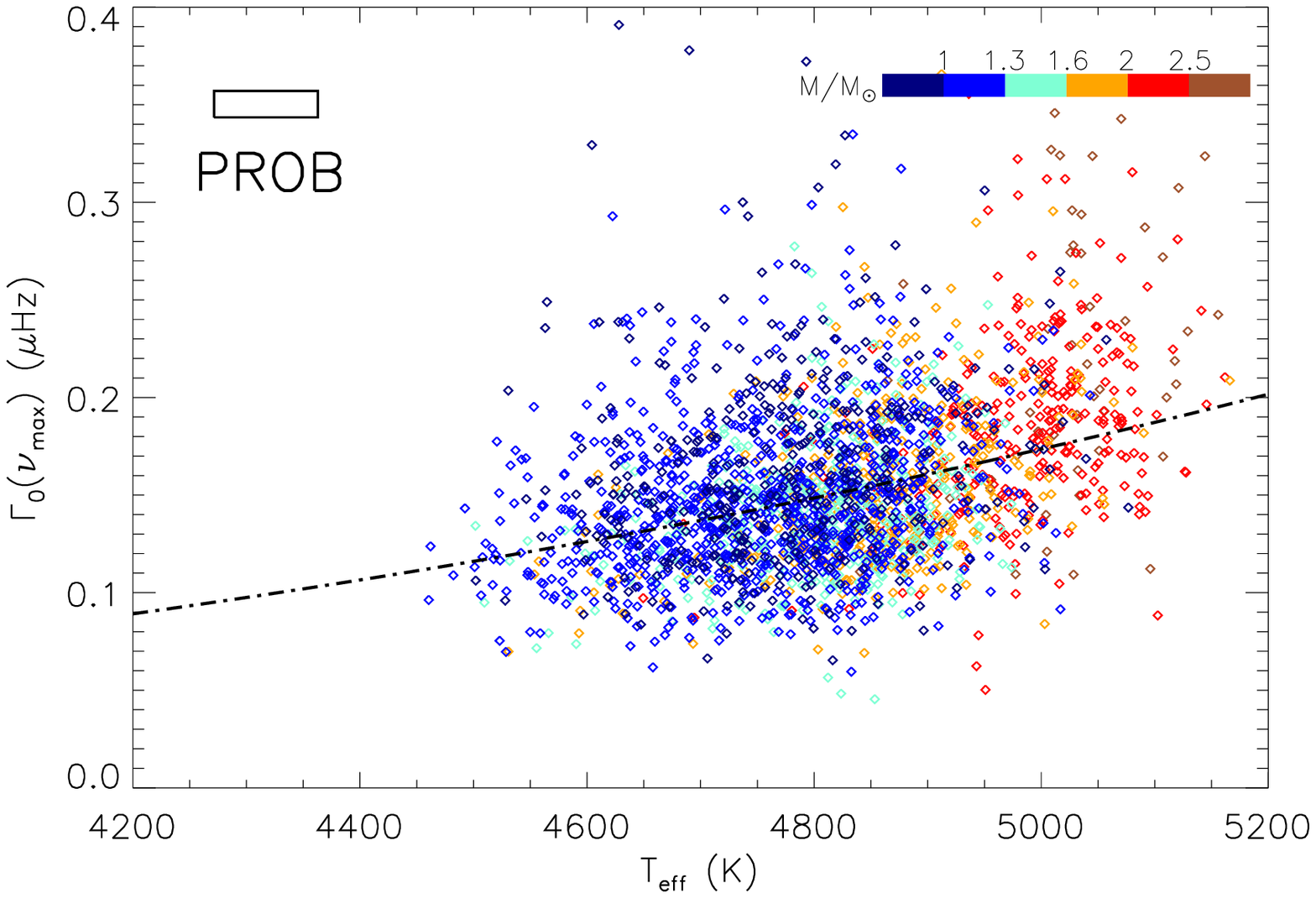}
  \includegraphics[width=9cm]{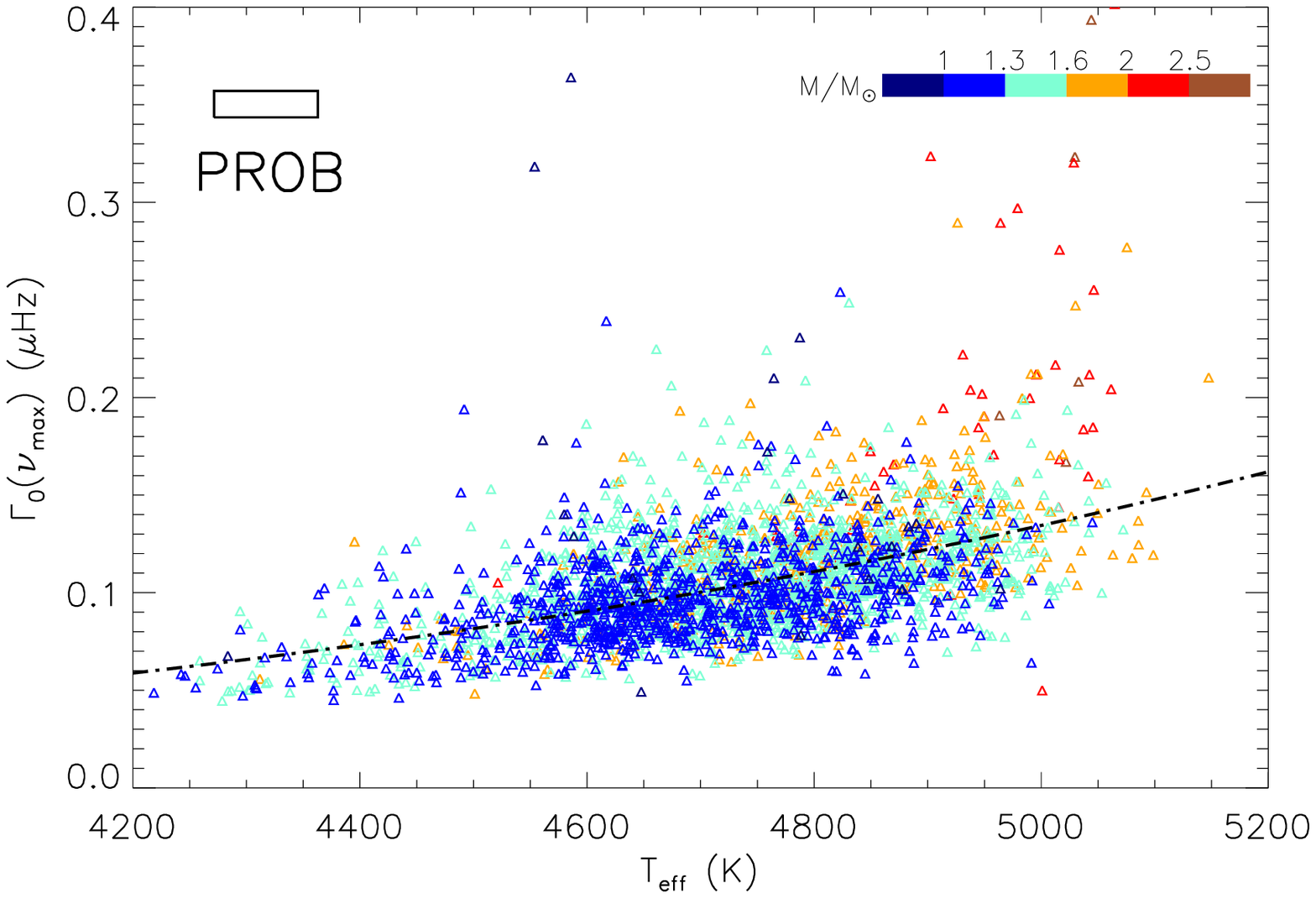}
  \caption{Global radial mode widths $\larg_n$ as a function of the effective temperatures $\Teff$ for the $\Vrard$ method (top) and for the $\Kallinger$ method (bottom). Clump and RGB stars are shown in the left and right panels, respectively, with their mass color-coded. The black squares indicate the typical uncertainties and the black dashed-dotted lines give the width-temperature relation fitted for each set of stars.}
  \label{fig:largeur_Teff_RGB}
\end{figure*}

\begin{table}[t]

\begin{center}
\caption{Temperature dependence of $\larg_n$ on the RGB for stars with different mass values} \label{table:Teff}
\begin{tabular}{ccccc}
 \hline
       $M/\Msol$     &    $1.0-1.2$  & $1.2-1.4$     & $1.4-1.6$     & $1.6-1.8$      \\
 \hline

 $\alphaT$($\Vrard$)         & $5.71\pm0.39$ & $5.03\pm0.45$ & $5.41\pm0.59$ & $5.78\pm0.69$  \\
 $\alphaT$($\Kallinger$)     & $4.55\pm0.27$ & $5.29\pm0.38$ & $4.54\pm0.40$ & $4.65\pm0.54$  \\

\hline
\end{tabular}
\end{center}

The temperature dependence of $\larg_n$ is modeled as $\larg_n \propto \Teff^{\alphaT}$.

\end{table}

The results are shown separately for RGB and clump stars in Fig.\,\ref{fig:largeur_Teff_RGB}, their values are consistent with the ones from previous studies \citep{2011A&A...529A..84B,2012ApJ...757..190C,2015A&A...579A..83C}. On the RGB, we note a double dependence: firstly, the radial mode width decreases with decreasing $\Teff$; secondly, at a given $\Teff$, the radial mode width increases with increasing mass. This shows that both the stellar mass and effective temperature have an influence on $\larg_n$, which was not reported in previous analyzes \citep[e.g.,][]{2011A&A...529A..84B}.

\begin{table}[t]

\begin{center}
\caption{Mass dependence of $\larg_n$ on the RGB for stars with different effective temperature values}\label{table:mass}
\begin{tabular}{cccc}
 \hline
       $\Teff$(K)    &   $4500-4700$    &   $4700-4900$   & $4900-5100$      \\
 \hline

 $\alphaM$($\Vrard$)         & $0.112\pm0.051$ & $0.138\pm0.036$ & $0.502\pm0.069$  \\
 $\alphaM$($\Kallinger$)     & $0.066\pm0.048$ & $0.251\pm0.034$ & $0.906\pm0.063$  \\

\hline
\end{tabular}
\end{center}

Mass dependence of $\larg_n$ on the RGB modeled as $\larg_n \propto (M/\Msol)^{\alphaM}$. 

\end{table}

In order to establish the degree of the dependence, we fit the linewidths as a function of $\Teff$. On the RGB, we find $\larg_n$ to vary with $\Teff^{5.28\pm0.17}$ for the mode widths obtained with the $\Vrard$ method and $\Teff^{4.74\pm0.13}$ for the ones obtained with the $\Kallinger$ method. In order to study the impact of the mass dependence on these fits, we perform several fits for stars with similar masses. The results can be seen on Table \ref{table:Teff}. They are consistent with each other.

The global mass dependence on the RGB can also be estimated with the same procedure. We find $\larg_n$ to scale with $M^{0.14\pm0.13}$ for the mode widths obtained with the $\Vrard$ method and with $M^{0.27\pm0.13}$ for the ones obtained with the $\Kallinger$ method. Here also, the temperature can have an influence on these measured dependencies. We then perform several fits for stars with similar temperature values to measure this impact. The results can be seen in Table \ref{table:mass}. The mass dependence is identified as being more important for stars with high $\Teff$ compared to stars with low $\Teff$. Both methods show the same behavior, the only difference is in the magnitude of the dependence that is higher for the $\Kallinger$ values at high $\Teff$. We also notice an important variation of the dependence as a function of the range of $\Teff$ selected. This phenomenon indicates a low reliability of the global mass dependence previously measured as shown by their large uncertainties. However, the variation of the results can also be caused by the low number of stars with high mass values present at low $\Teff$ leading to less accurate fits. These results show that a precise characterization of the mass dependence independent of the influence of the temperature is currently difficult for RGB stars and that the performed measurements have to be considered as an approximation. This problem could be overcome by the use of population synthesis in order to identify and correct all possible biases but this is beyond the scope of this paper.
\newline

For clump stars, we find $\larg_n$ to vary with $\Teff^{4.10\pm0.17}$ for the $\Vrard$ method and $\Teff^{3.82\pm0.21}$ for the $\Kallinger$ method, which correspond to lower exponents than for RGB stars. This can be explained by the fact that clump stars with similar masses have similar effective temperatures, as it can be seen on Fig. \ref{fig:largeur_Teff_RGB}. The presence of high-mass stars with high $\Teff$ will then artificially modify the measured dependence in that specific case. In fact, the absence of important $\Teff$ evolution during the clump phase on the contrary to the RGB does not readily allow us to disentangle the mass and temperature dependencies.
\newline

The exponent of the observed global power laws turns out to be much smaller for red giants than for less-evolved stars, with exponents between $15.3\pm1.9$ \citep{2017ApJ...835..172L} and $16.7\pm1.8$ \citep{2012A&A...537A.134A}. Moreover, the observed relationship seems to differ from theoretical expectations. The small exponent we find for the power law does not appear to correspond to the high value predicted by theory \citep[10.8 following][]{2012A&A...540L...7B}. However, previous theoretical works as the one performed by \citet{2012A&A...540L...7B} took main-sequence stars as well as red giants into account to deduce the $\larg_n$-$\Teff$ relation. It is thus not relevant to compare the theory to the exponent we measured based on red giants only.

Although our sample covers a limited range in effective temperature, so that any possible bias in $\Teff$ may affect the fit, our results are based on the largest sample of red giants analyzed so far when addressing radial mode widths and, consequently, we may consider them to benefit from a high statistical reliability.
\newline

Finally, we investigate the impact of the stellar metallicity on the mode width but find no significant influence (see Appendix \ref{scaling_largeur_Teff_met})

\subsection{Individual mode width variation with frequency}

The variation of the individual mode widths ($\largeurr$) with frequency for a single star was previously analyzed for subgiants and main-sequence solar-type pulsators \citep{2014A&A...566A..20A,2017ApJ...835..172L}. This variation exhibits a similar behavior for all solar-type pulsators with low effective temperature. It is characterized by a global increase of the mode width with increasing frequency along with a small $\largeurr$ dip around $\numax$ \citep[e.g., ][]{2017ApJ...835..172L}. The origin of the dip was theoretically investigated by \citet{2012A&A...540L...7B} who attributed it to the competition between the perturbation of the turbulent pressure and the perturbation of entropy. These two physical mechanisms have opposite impacts: one contributes to the damping and the other to the driving, so that they compensate each other. The maximum compensation between the two terms happens around $\numax$, where it induces a dip in the global increase of the mode width. Following that study, the $\largeurr$ dip should be observed for the mode widths in all solar-like pulsators, including red giant stars. This was already pointed out by \citet{2015A&A...579A..83C} for most of the $19$ low-RGB stars they analyzed. The purpose of this section is to see whether this behavior can be observed in a larger sample.
\newline

\begin{figure*}[t]              
  \includegraphics[width=9cm]{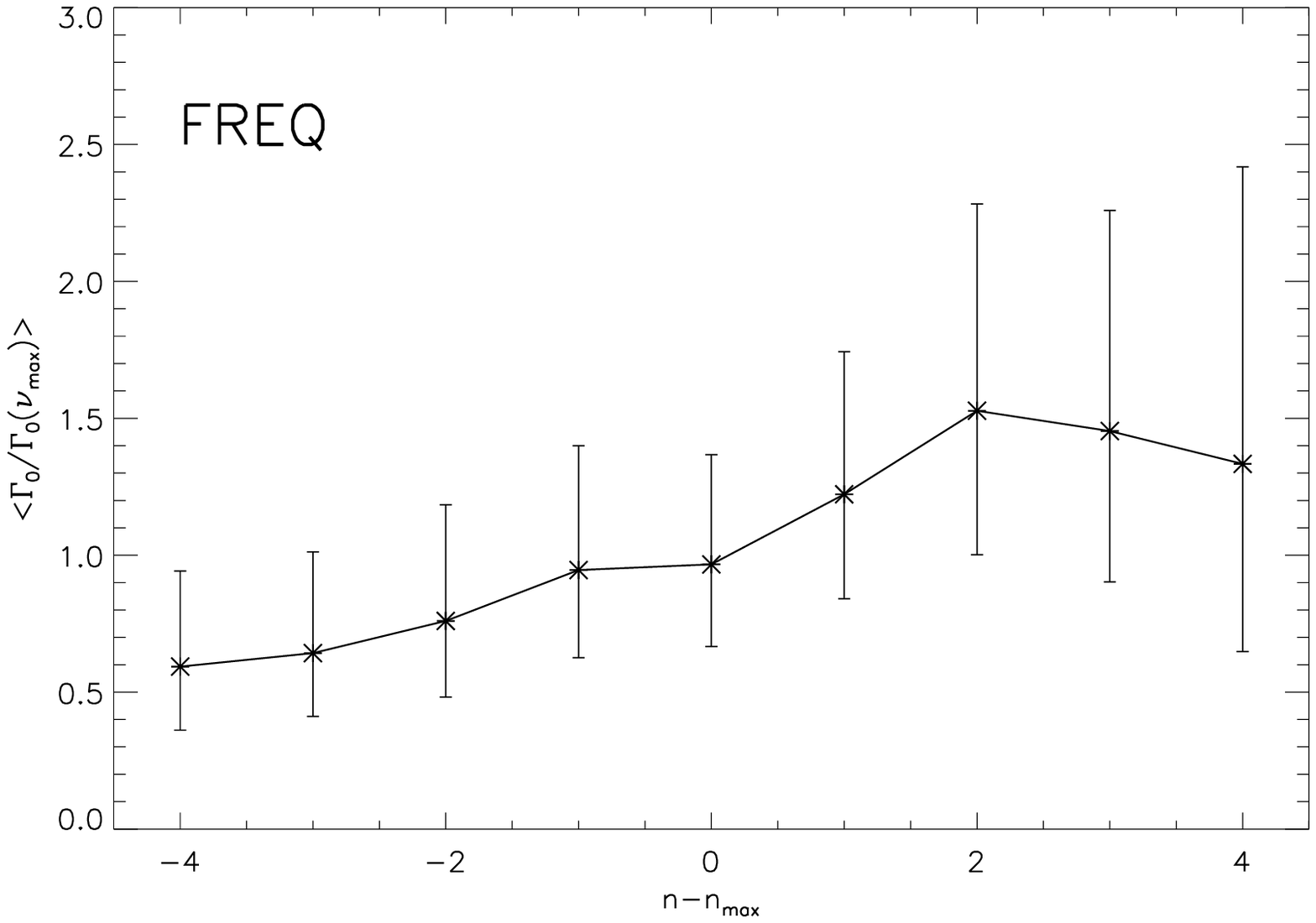}
  \includegraphics[width=9cm]{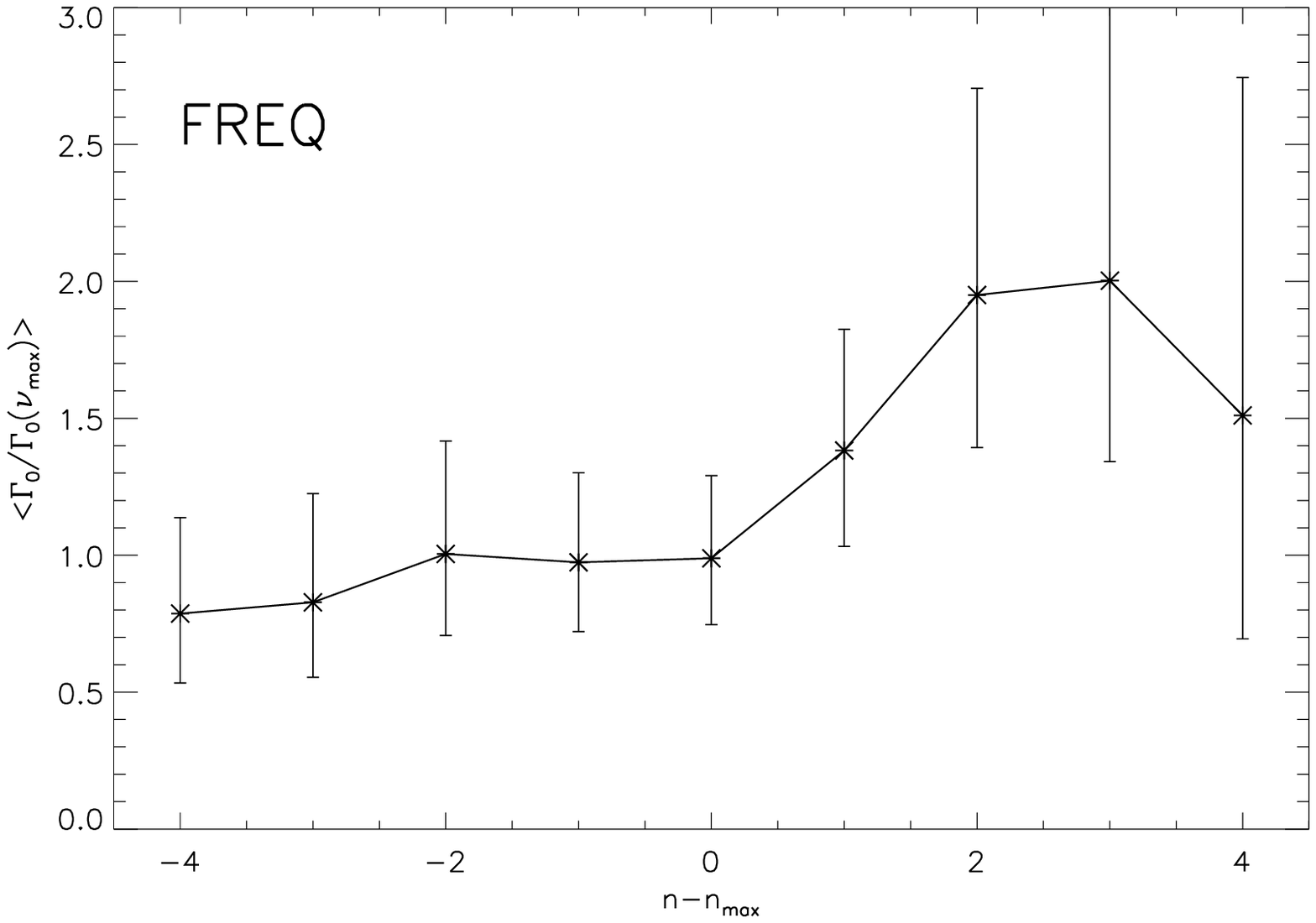}
  \includegraphics[width=9cm]{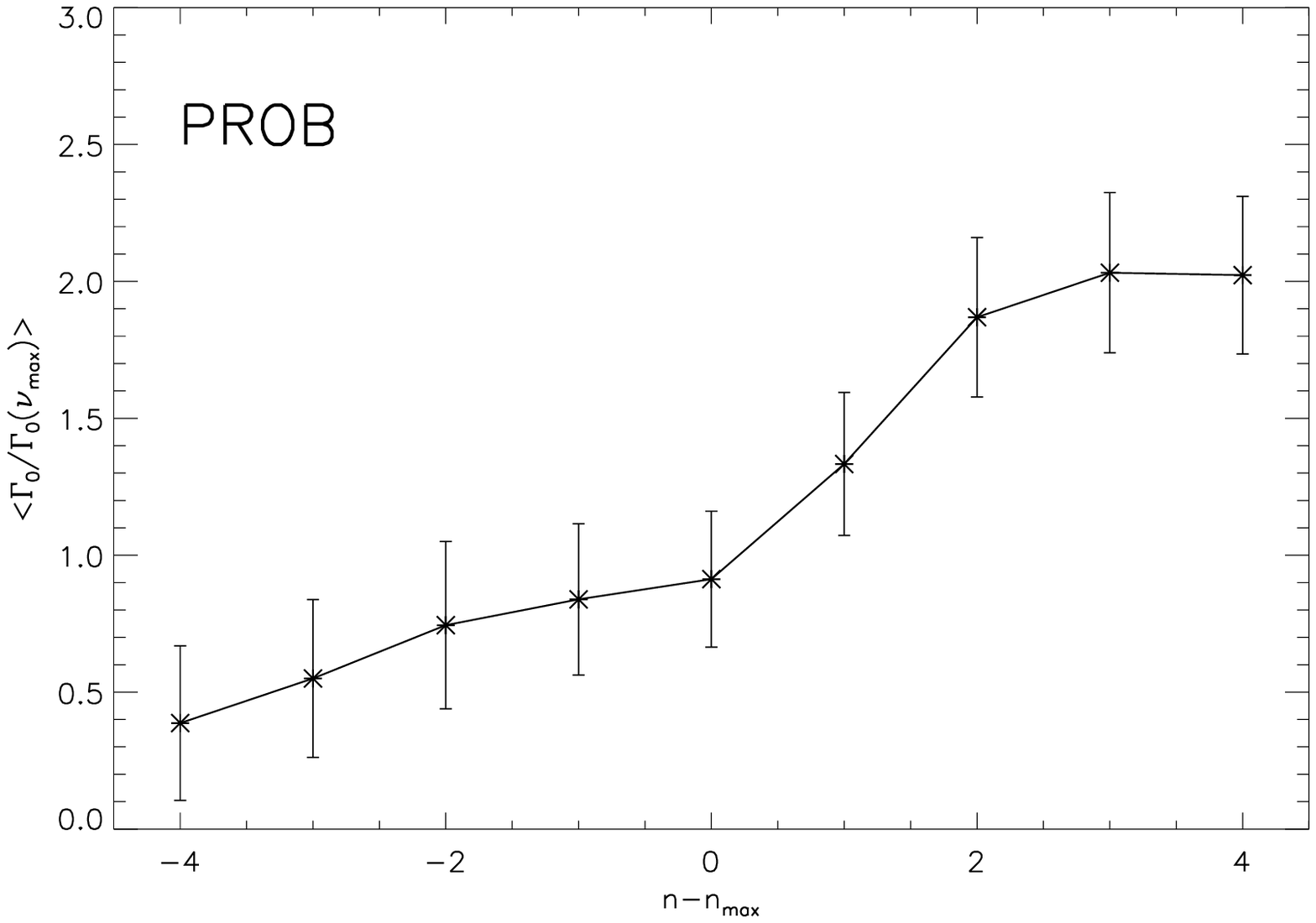}
  \includegraphics[width=9cm]{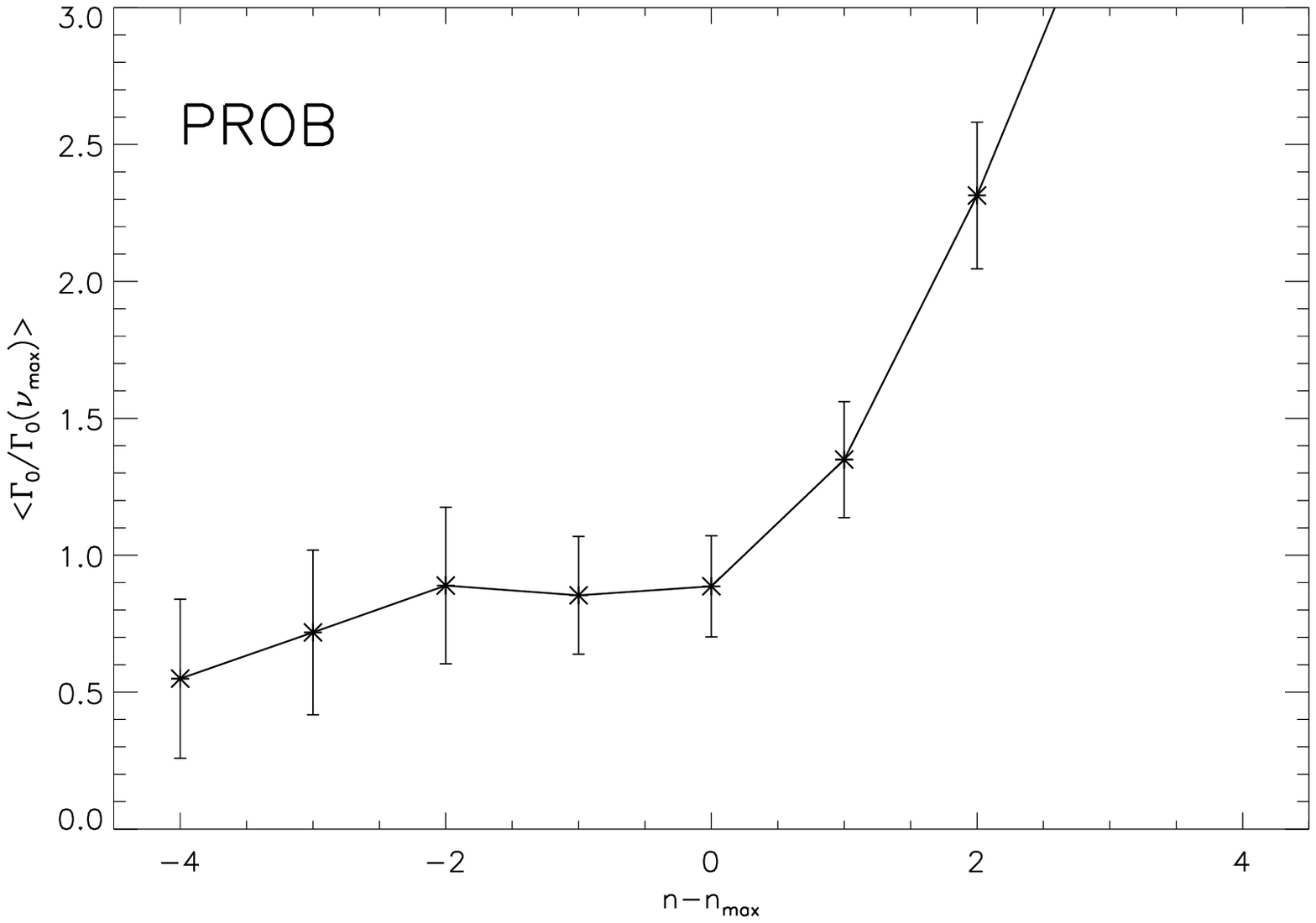}
  \caption{Collapsogram of the relative radial mode width as a function of the radial mode order $n-\nmax$ for the $\Vrard$ method (top) and for the $\Kallinger$ method (bottom). Clump and RGB stars are shown in the left and right panels, respectively. The error bars correspond to the $1\sigma$ uncertainties.}
  \label{fig:collapse}
\end{figure*}

We estimate the global shape of the $\largeurr$ variation with frequency by using a collapsogram on the mode radial order. It corresponds to the calculation of the mean ratio between $\largeurr$ and $\larg_n$ as a function of the radial mode order centered on $\numax$. Firstly, we identify the radial order $n$ of each radial mode and the value of $n$ for the radial mode closest to the maximum oscillation power. This order is called $\nmax$ and is obtained with the following relation $\nmax = \Dnu/\numax - \varepsilon$, where $\varepsilon$ is the offset of the asymptotic pattern. Secondly, we calculate the ratio between the individual mode widths and $\larg_n$. Finally, we compute collapsograms by taking the mean value of these ratios with respect to the relative radial order $n-n_\mathrm{max}$. This was done for clump and RGB stars separately, following the differences that arise between the two evolutionary states. The results are shown in Fig.\,\ref{fig:collapse}.
\newline

Both methods show an increase of the relative radial mode width with a slow-down around $\nmax$ for RGB stars only, which could correspond to the $\largeurr$ dip observed for main-sequence and subgiants solar-like pulsators even if the large uncertainties do not allow firm conclusions. For the $\Vrard$ method, we can also note an apparent stabilization or even a decrease of the relative radial mode width at high radial orders (above $\nmax+2$), which is not found in other solar-like pulsators. However, the large uncertainties for the higher radial orders and the low number of stars for which we obtain a radial mode width measurement at these radial orders question the physical meaningfulness of this observation. In fact, this behavior is not observed for the mode width obtained with the $\Kallinger$ method. Instead, we note a continuous increase of the mode width, casting doubts about the reality of the behavior of the relative radial mode width at high radial order for the $\Vrard$ method.

As mentioned before, the increase in the relative radial mode width undergoes a slow-down near $\nmax$ for RGB stars in both methods. However, this behavior does not appear for the clump stars in the $\Vrard$ method and is significantly reduced in the $\Kallinger$ method. This could indicate that the dominant physical mechanism behind mode damping changes between the RGB and clump evolutionary states. However, the large uncertainties prevent us from drawing a firm conclusion and further studies are needed in order to confirm this behavior.

Finally, we check the influence of the stellar mass, surface gravity and $\Teff$ on the result. We find no significant differences when restricting the collapsogram to either low-mass stars or high-mass stars. Therefore, we conclude that the mass influence on these results is negligible. When restricting the collapsograms to low-$\Teff$ stars or stars with low surface gravity, the dip appears more distinctly but, considering the error bars, no firm conclusions could be drawn.
\newline

\section{Mode amplitudes}\label{sec:mode_amplitude}

In this Section, we discuss the results we obtain for the relative amplitudes and compare them with theoretical predictions and previous observational results.

\subsection{Amplitude variation with stellar evolution}

\begin{figure}[t]              
  \includegraphics[width=9cm]{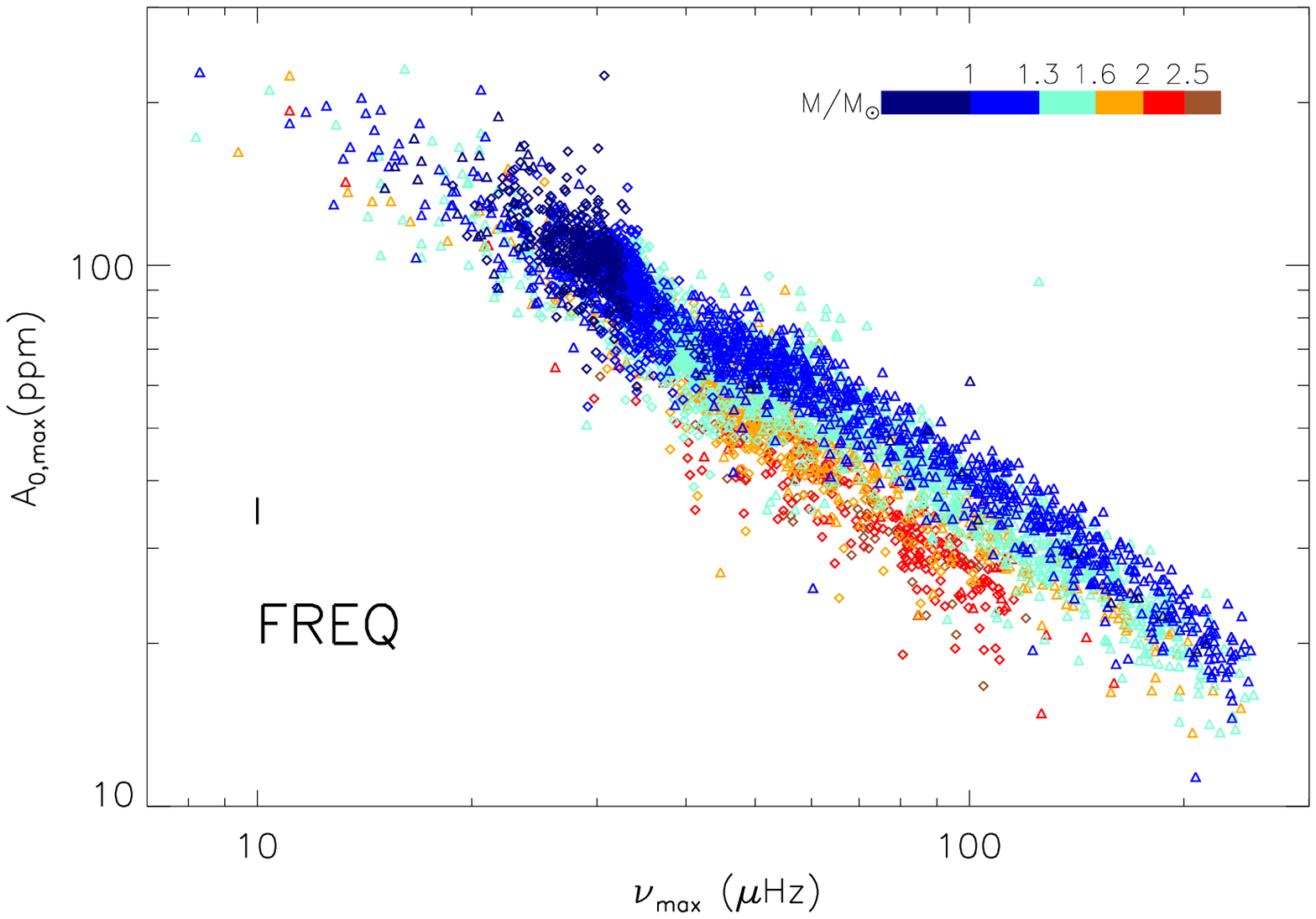}
  \includegraphics[width=9cm]{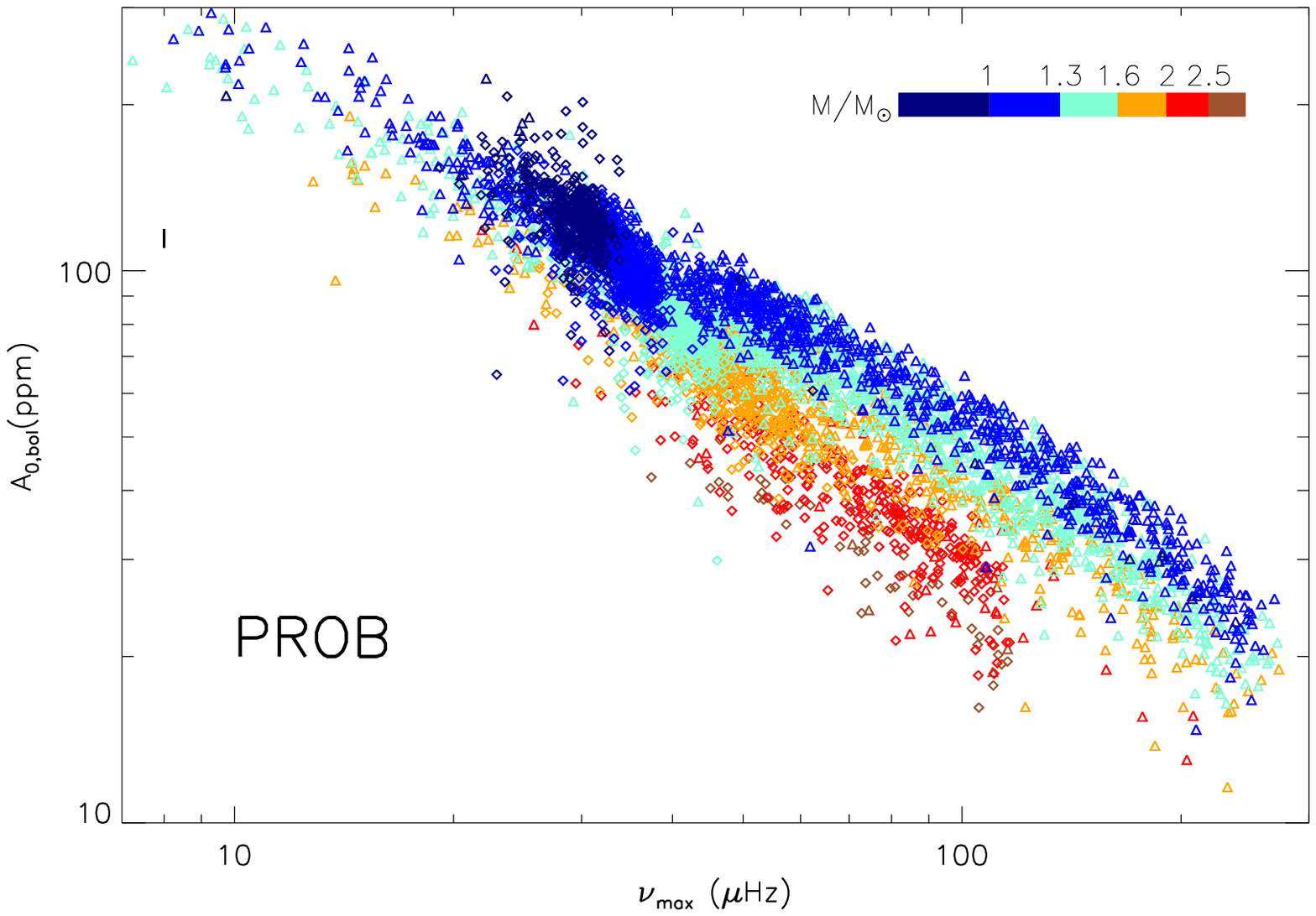}
  \caption{Bolometric amplitudes $\ampbol$ as a function of $\numax$ for the $\Vrard$ method (top) and for the $\Kallinger$ method (bottom). The color code indicates stellar mass. The diamond and triangle symbols indicate clump and RGB stars, respectively. The black line corresponds to the mean uncertainties.}
  \label{fig:amplitude_gene}
\end{figure}

The variation of $\ampbol$ with $\numax$ is shown in Fig.\,\ref{fig:amplitude_gene} for both methods. As for the mode widths, the maximum mode amplitudes obtained with the two methods show a similar behavior. The decrease of the stellar envelope density with stellar luminosity is manifested by the increase of $\ampbol$ with decreasing $\numax$. This decrease in envelope density enables a more vigorous convection and, consequently, a more efficient mode excitation. We also note, for both samples and both evolutionary states, a clear mass dependence: stars with a higher mass have a lower maximum mode amplitude, as previously noticed \citep{2010ApJ...723.1607H,2011A&A...532A..86M,2011ApJ...737L..10S,2011ApJ...743..143H,2012A&A...537A..30M,2014A&A...570A..41K}. However, we can clearly see that the mass dependence is present in an equivalent manner for both evolutionary states.

On the contrary to the global behavior, the measured amplitude values differ between the two methods. For RGB stars with low $\numax$, the $\Vrard$ amplitudes are globally underestimated by about $20\%$ with respect to the $\Kallinger$ values. On the other hand, clump stars are less affected, leading to a small overestimation of the amplitudes we obtained with the $\Vrard$ method compared to the $\Kallinger$ method for stars with high $\numax$. The origin of this difference is not known.

One last thing to observe is the similarity between clump and RGB star amplitudes. At similar mass and seismic parameters, stars in both evolutionary states have approximately the same maximum amplitude (Fig.\,\ref{fig:amplitude_gene}).

\subsection{The amplitude\textbf{--}$L/M$ relation} \label{sec:ampLM}

Following \citet{1995A&A...293...87K,1999A&A...351..582H,2011A&A...529L...8K} and later \citet{2012A&A...543A.120S}, the bolometric amplitude of solar-like oscillations is expected to scale as
\begin{equation}\label{eqt-scaling-A}
  \ampbol \propto \tau\ind{max}^{1/2} \ \left({L\over M}\right)^{a} \ \Dnu^b 
  .
\end{equation}  
In order to test this theoretical scaling we need an estimate for $L$, which is provided by the scaling relations of $\numax \propto M/R^2/\sqrt{\Teff}$ \citep[e.g.,][]{1991ASPC...20..139B,1995A&A...293...87K} and $L \propto R^2\Teff^4$. It follows \citep[e.g.,][]{2011A&A...529A..84B},
\begin{equation}
   L \propto M \Teff^{7/2} \numax^{-1} 
    .
   \label{Equation:luminosity-Teff}
\end{equation}
Since $\Dnu$ is expected to scale as $\Dnu \propto \sqrt{M/R^3}$ \citep[e.g.,][]{1986ApJ...306L..37U,1994ARA&A..32...37B} we can substitute $\Dnu$ in Eq. (\ref{eqt-scaling-A}). Assuming $\larg_n \propto M^c \Teff^d$ we can then write,
\begin{equation}\label{eqt-scaling-A2}
  \ampbol \propto L^{a-3b/4} \ M^{-a+b/2-c/2} \ \Teff^{-d/2 + 3b}
  .
\end{equation}
From this development, it can clearly be seen that the $b$ exponent, which represents the $\Dnu$ dependence (Eq.~\ref{eqt-scaling-A}), does not produce an important influence in the departure from the $L/M$ power law. On the contrary, the influence of the $c$ exponent, introduced by $\Gamma$, amplifies the departure from a power law in $L/M$.

In order to simplify the approach, we consider that the relative variation of the effective temperature on the RGB and in the clump is so small compared to the variation in luminosity that we can neglect the $\Teff$ dependence. Therefore, we performed a fit of $\ampbol$ as power law of $L$ and $M$ only. We find a variation in $L^{0.77\pm0.06} M^{-1.25\pm0.14}$ with the $\Vrard$ method, and in $L^{0.70\pm0.06} M^{-1.40\pm0.13}$ with the $\Kallinger$ method. In order to evaluate the validity of the previous hypothesis leading to neglect the $\Teff$ influence, the variation of the mode amplitude with $L$, $M$ and $\Teff$ was also measured. We find a variation in $L^{0.70\pm0.07} M^{-1.03\pm0.15} \Teff^{-4.79\pm1.36}$ with the $\Vrard$ method, and in $L^{0.61\pm0.06} M^{-1.14\pm0.14} \Teff^{-5.48\pm1.18}$ with the $\Kallinger$ method. The dependence in $L$ and $M$ are less important, as expected, but consistent with the ones we previously found, considering the error bars. In contrast to these results, we can see that the measured $\Teff$ dependence is very important. However, the large error bars on the exponent and the lack of important impact on the other dependencies lead us to confirm the simplification hypothesis which correspond to neglect the $\Teff$ dependence. 

To assess the impact of the clump stars on the fits, we also consider the RGB stars alone and find significant differences: $\ampbol$ varying as $L^{0.84\pm0.08} M^{-1.34\pm0.27}$ with the $\Vrard$ method, and in $L^{0.82\pm0.07} M^{-1.61\pm0.27}$ with the $\Kallinger$ method. As stated above, seismic and stellar parameters of stars do not evolve much during the helium burning phase, on the contrary to the RGB phase. It follows that clump stars with similar masses will have similar $\ampbol$ values thus making it difficult to disentangle the luminosity and mass dependence for these stars. The fit performed only on RGB stars allows us to get rid of a possible influence of one dependence on the other, therefore, it will be the one considered throughout the rest of the article.

The exponents we find for the luminosity are slightly larger than the ones theoretically predicted \citep{2012A&A...543A.120S}, but agree with previous measurements \citep{2010A&A...517A..22M,2011ApJ...743..143H,2011ApJ...737L..10S,2012A&A...537A..30M,2013MNRAS.430.2313C,2014A&A...570A..41K}. A possible explanation for the difference between the predicted and observed exponents is the strong influence of non-adiabatic effects on the mode driving in red giants. Indeed, the theoretical predictions previously mentioned do not rely on fully non-adiabatic models. This type of modeling has not yet been performed \citep{2013EPJWC..4303008S}.
\newline

The observed exponents in our fits help constraining the theoretical predictions. We derive the exponents $a = 2.13\pm1.16$ and $b = 1.72\pm1.66$ for the $\Vrard$ method and $a = 2.77\pm1.14$ and $b = 2.60\pm1.62$ for the $\Kallinger$ method. We note that the relationship between the mode width $\larg_n$ and the stellar mass helps understanding the different exponents in $L$ and $M^{-1}$. However, these exponents are significantly larger than the ones predicted by theoretical works, corresponding to $a = 1.55\pm0.09$ and $b = 0.50\pm0.03$ \citep{2012A&A...543A.120S}. Even if large uncertainties are present in our results, the discrepancies between observations and theoretical works show that more studies are needed in order to produce realistic models describing the physical mechanisms, which govern mode excitation.

\subsection{Evidence of a metallicity influence}

\begin{figure*}[t]              
  \includegraphics[width=9cm]{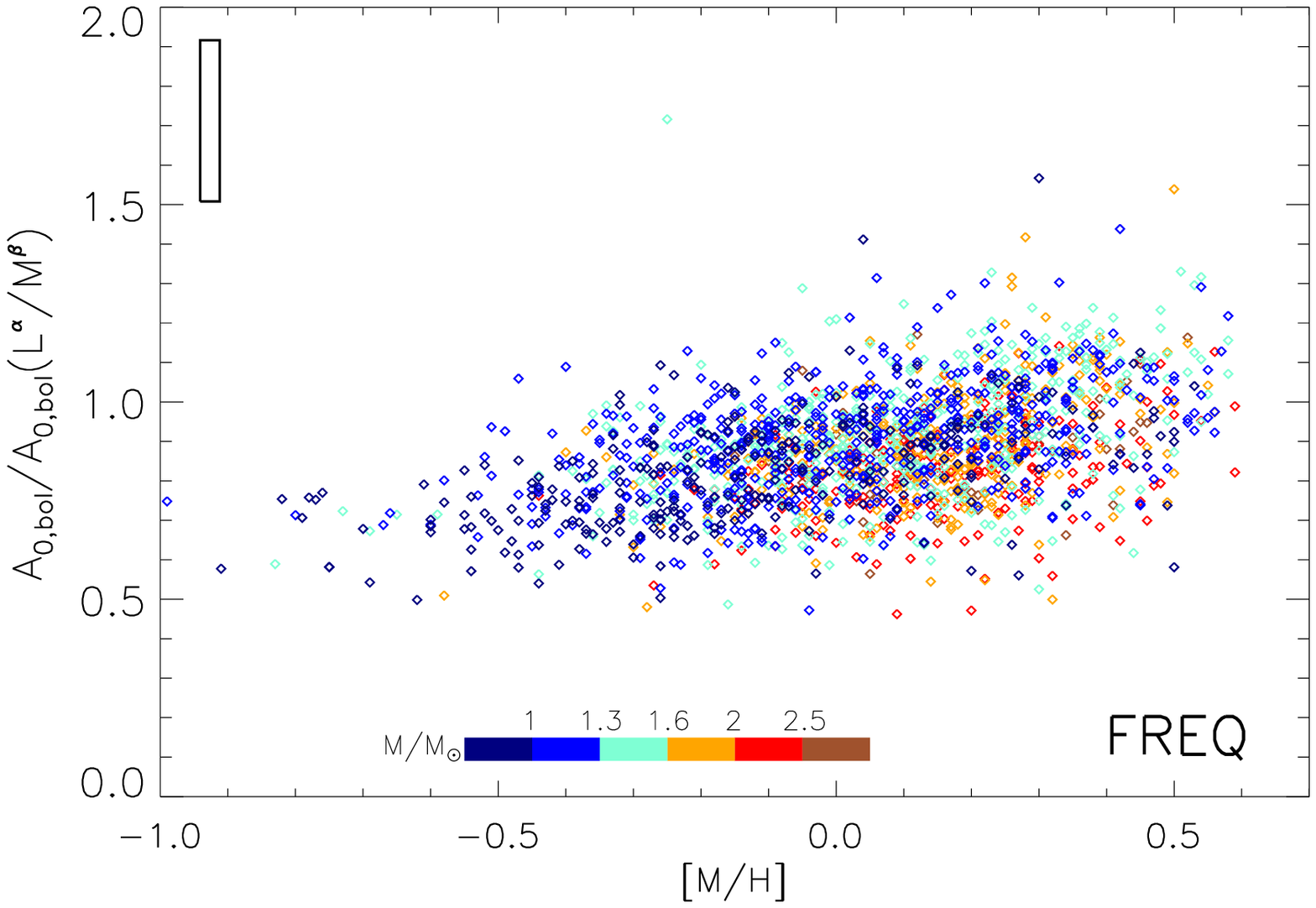}
  \includegraphics[width=9cm]{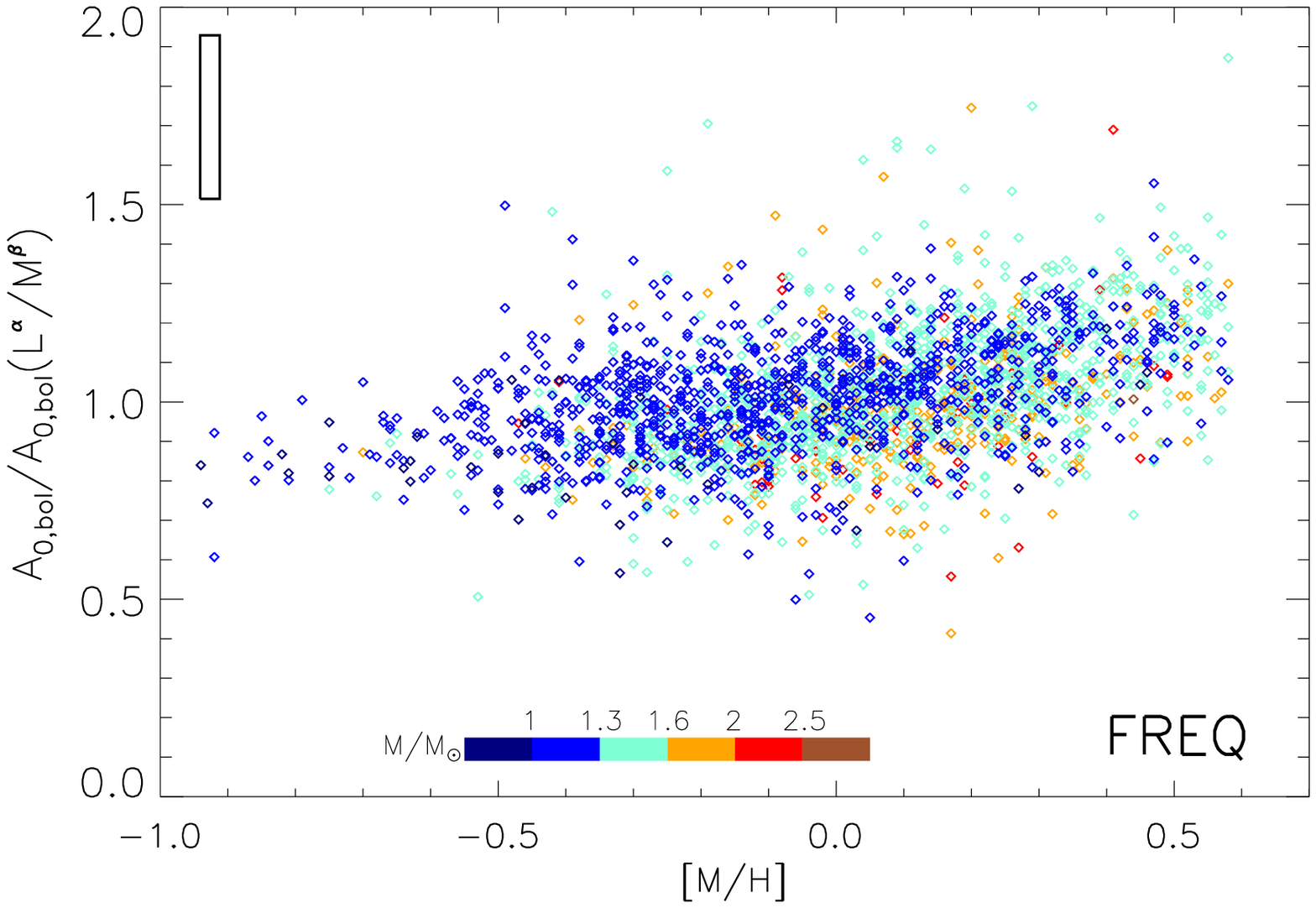}
  \includegraphics[width=9cm]{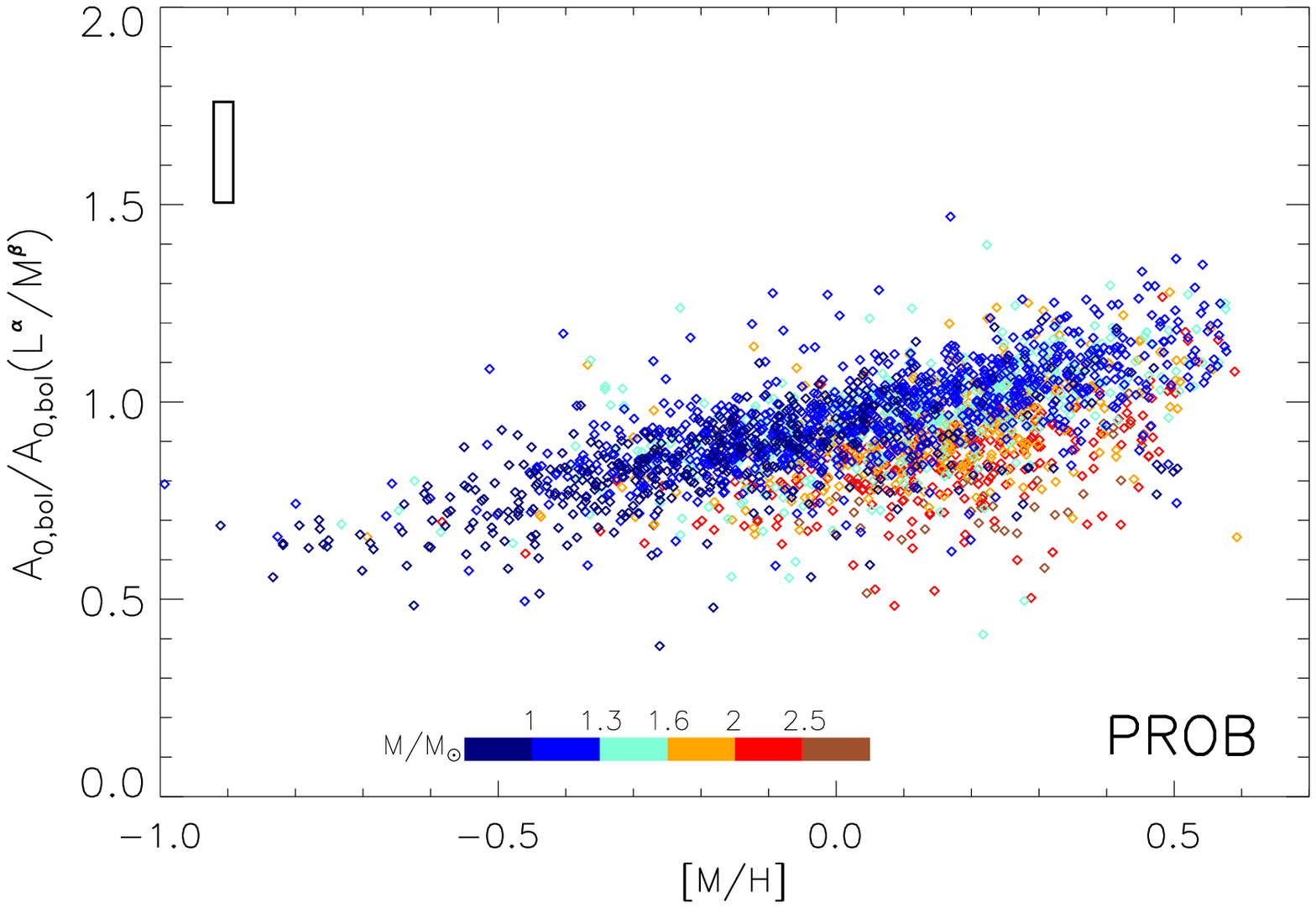}
  \includegraphics[width=9cm]{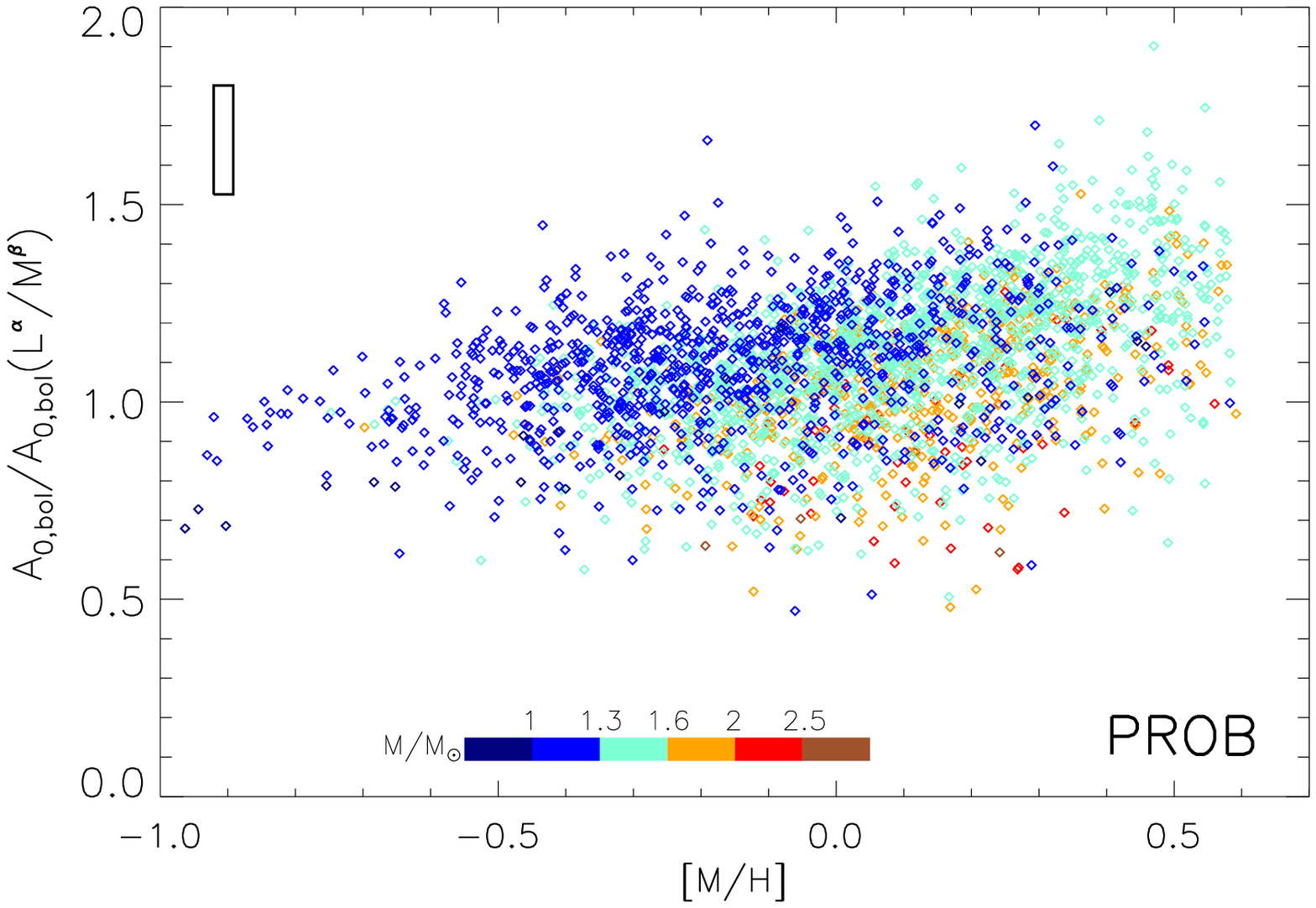}
  \caption{Bolometric amplitudes ($\ampbol$) corrected for the $L$ and $M$ dependencies as a function of the stellar metallicity for the $\Vrard$ method (top) and the $\Kallinger$ method (bottom), with the stellar mass color-coded. Clump stars are shown in the left panels and RGB stars in the right panels. The black rectangle corresponds to the typical uncertainties.}
  \label{fig:amplitude_metal}
\end{figure*}

Several theoretical studies suggest that stellar metallicity has an influence on the mode amplitude \citep{1999A&A...351..582H,2010A&A...509A..15S}. In order to test this, we plot the mode amplitudes,  cleared of the other dependencies we caracterized previously, as a function of the stellar metallicity (Fig.\,\ref{fig:amplitude_metal}). The metallicities are taken from the APOKASC catalog, providing [M/H] measurements for $4367$ objects from which a $\ampbol$ and $\larg_n$ value could be measured with the $\Vrard$ method and for $5339$ objects from the $\Kallinger$ method.

The identification of a metallicity dependence requires the suppression of the luminosity and mass dependences previously identified. We therefore use the fits from Sec.\,\ref{sec:ampLM} to correct the amplitudes for the influence of $L$ and $M$ previously characterized. We estimate the dependence of the normalized amplitudes on metallicity separately for clump and RGB stars in order to remove a potential influence from the stellar evolutionary status. We find a clear increase of the mode amplitudes with increasing metallicity for both evolutionary states and both methods in agreement with the previous analysis \citep{2012A&A...537A..30M}. However, we find a slightly higher dependence in our data: an increase of one dex in metallicity causes an increase of more than $20\%$ in amplitude. More precisely, for the $\Vrard$ method, the one dex increase corresponds to a $25\%\pm3\%$ increase in amplitude for RGB stars and $27\%\pm3\%$ for clump stars. For the $\Kallinger$ method, the rise corresponds to $45\%\pm2\%$ for RGB stars and $49\%\pm2\%$ for clump stars, which is slightly higher than the results we obtain with the other method.

The overall results are in agreement with theoretical predictions \citep[e.g.,][]{1999A&A...351..582H,2010A&A...509A..16S,2010A&A...509A..15S}, which suggest larger pressure mode amplitudes for metal-rich stars than for metal-poor stars. Even though we still have to investigate if our observational results are comparable to these predictions in a more quantitative way, such an agreement indicates that the global understanding of the variation of the mode amplitude with the stellar physical parameters is already quite good.

These results are also in agreement with the metallicity influence observed for the granulation which correspond to an increase of the granulation amplitude with the stellar metallicity \citep[e.g.][]{2017A&A...605A...3C}, since mode and granulation amplitude are linked together \citep{2011A&A...529L...8K}. However, it appears that the metallicity influence is much weaker for the mode amplitude compared to the one found by \citet{2017A&A...605A...3C} for granulation amplitude.

\section{Conclusion}

In this study, we measure the radial mode widths and amplitudes in thousands of red giant stars. Our results emphasize a relationship between the radial mode width and the effective temperature, which can be modeled by a power law. The exponent of the power law is, however, much smaller than those previously observed for subgiants and main-sequence stars but is consistent with previous measurements done with red giants. We also find a significant mass dependence of the radial mode width and that the mode widths are sensitive to the evolutionary state of a star, with clumps stars generally having broader radial modes than RGB stars as previously shown by \citet{2012ApJ...757..190C}. The mode widths are directly related to mode damping so that their measurements offer new constraints on this parameter. Since our results are not completely compatible with theoretical predictions, they suggest that the contributions of the different physical mechanisms to mode damping in red giants have to be specified and work in a different manner than in main-sequence stars. Further theoretical studies will be necessary to fully exploit the new constraints we find for this parameter.

On the contrary, our results for the mode amplitudes confirm and extend previous ones, in particular on the influence of the mass and the relation between the mode amplitudes and the luminosity-to-mass ratio. We can also confirm a clear influence of the stellar metallicity on the mode amplitudes: an increase in metallicity causes an increase of the mode amplitude. The theoretical prediction of such a behavior seems to indicate that the physical mechanism behind mode excitation is already well understood. However, differences between the observed dependence (especially the dependence with the stellar mass and luminosity) and theoretical predictions show that some further studies are necessary in order to develop realistic models taking all non-adiabatic effects into account.

\begin{acknowledgements}
MV acknowledges funding by the Portuguese Science and Technology Foundation (FCT) through the grant with reference CIAAUP-03/2016-BPD, in the context of the project FIS/04434, co-funded by FEDER through the program COMPETE.

MSC was supported by FCT through national funds (Investigador contract of reference IF/00894/2012 and UID/FIS/04434/2013) and by FEDER through COMPETE2020 (POCI-01-0145-FEDER-030389 and POCI-01-0145- FEDER-007672).
\end{acknowledgements}

\bibliographystyle{aa} 
\bibliography{largeur_BM}

\appendix

\section{Precise mode-fitting results for each individual star}\label{Section: CDS_data_individualstars}

The $\Vrard$ method allows us to realize precise mode fitting of several radial modes in the oscillation spectra of each of the $5221$ stars that were analyzed in this study. The results are available in the form of several files (one for each star) at the CDS. An example of the shape of the different files is given on Table \ref{table: CDS_individual} for the star KIC1027337

\begin{table*}[t]

\begin{center}
\begin{tabular}{cccccccc}
 \hline
Frequency  &  $\delta$Frequency  & $\largeur_g$ & $\delta_{U}\largeur_g$ & $\delta_{L}\largeur_g$ &  Amplitude  &  $\delta_{U}$Amplitude  &  $\delta_{L}$Amplitude \\
($\mu$Hz) & ($\mu$Hz) & ($\mu$Hz) & ($\mu$Hz) & ($\mu$Hz) & (ppm) & (ppm) & (ppm) \\
 \hline

56.94  & 0.0136 & 0.0815 & 0.0413 & 0.0274 &  25.76  & 14.079 & 9.090  \\
63.51  & 0.0091 & 0.0702 & 0.0190 & 0.0149 &  34.37  & 10.625 & 8.080  \\
70.47  & 0.0086 & 0.1463 & 0.0186 & 0.0165 &  53.81  &  7.200 & 6.348  \\
77.39  & 0.0105 & 0.1280 & 0.0342 & 0.0270 &  68.25  & 21.831 &16.461  \\
84.36  & 0.0142 & 0.1371 & 0.0367 & 0.0290 &  40.13  & 12.597 & 9.557  \\
91.65  & 0.0305 & 0.1649 & 0.0957 & 0.0606 &  18.46  & 10.701 & 6.777  \\

\hline
\end{tabular}
\end{center}

\caption{Radial mode parameters for the star KIC1027337. Tables for the $5221$ stars that were used in this study is available at the CDS. The columns correspond to, from left to right, the mode frequencies, the $1\sigma$ uncertainties on this parameter, the mode widths ($\largeur_g$), the upper $1\sigma$ uncertainties on this parameter, its lower $1\sigma$ uncertainties, the mode amplitudes, the upper $1\sigma$ uncertainties on this parameter and its lower $1\sigma$ uncertainties. } \label{table: CDS_individual}

\end{table*}

\section{Metallicity influence on mode width values}\label{scaling_largeur_Teff_met}

\begin{figure}[t]              
  \includegraphics[width=9cm]{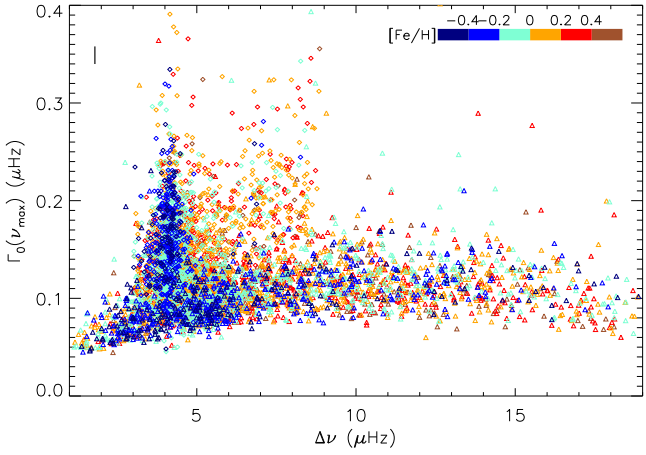}
  \caption{Global radial mode width ($\larg_n$) as a function of the large separation ($\Dnu$) for the $\Kallinger$ method, with the stellar metallicity color-coded. The diamond and triangle symbols indicate clump and RGB stars, respectively. The mean uncertainty in $\larg_n$ is given by the vertical black line in the upper left corner.}
  \label{fig:largeur_met}
\end{figure}

The influence of the metallicity on the mode width was previously mentioned as a possibility (Baudin, private communication). We, then, investigated it in this study. Fig. \ref{fig:largeur_met} shows the variation of $\larg_n$ as a function of the large separation for the $\Kallinger$ method and with a color code indicating the star metallicities. The metallicities are taken from the APOKASC catalog, providing [M/H] measurements for $4367$ objects from which a $\ampbol$ and $\larg_n$ value could be measured with the $\Vrard$ method and for $5339$ objects from the $\Kallinger$ method. The secondary clump stars exhibit higher metallicity values as expected for younger stars but no other trend is observed for the results from both methods.

\begin{figure}[t]              
  \includegraphics[width=9cm]{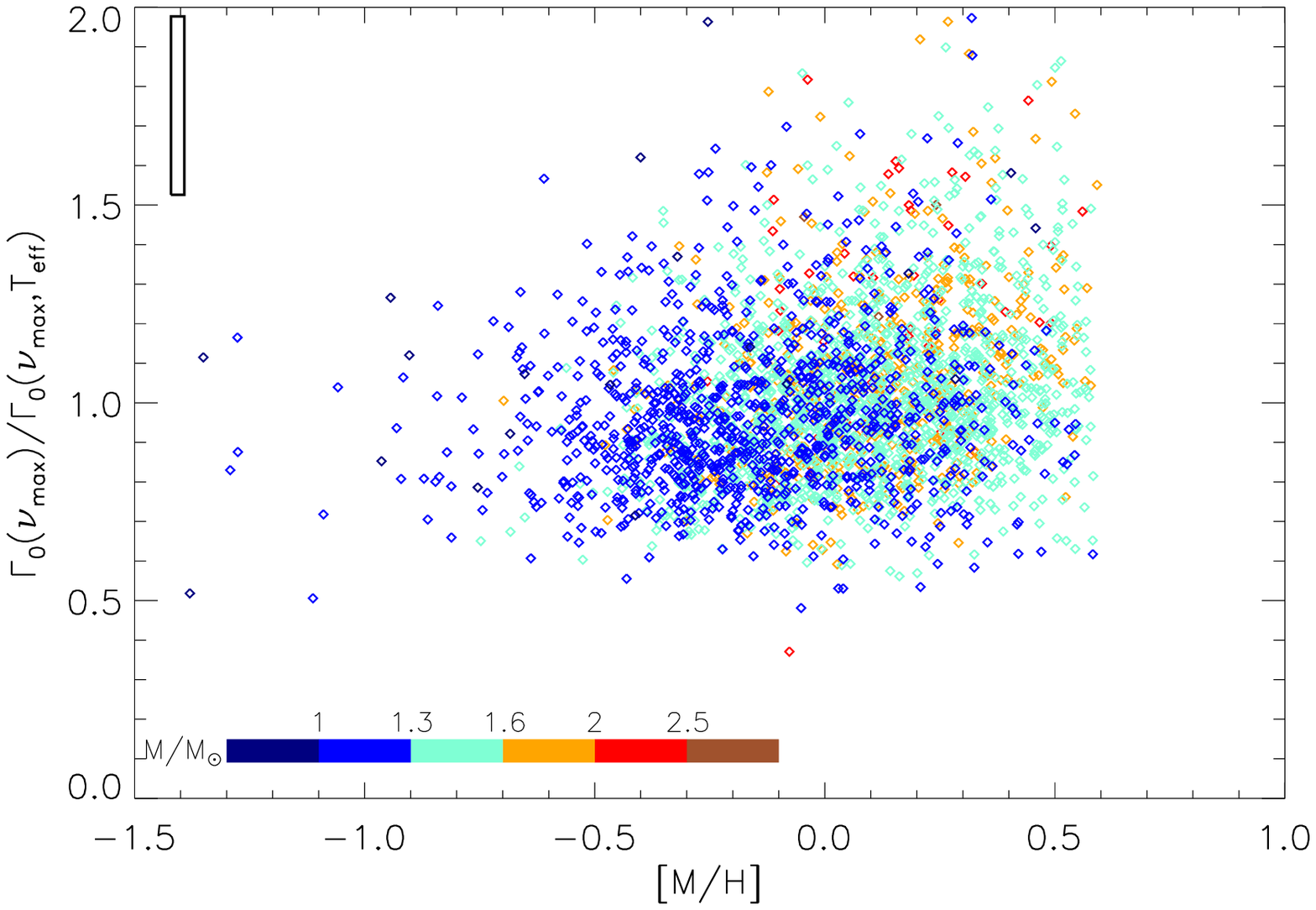}
  \caption{Global radial mode width ($\larg_n$) of RGB stars corrected from the $\Teff$ dependence as a function of the metallicity for the $\Kallinger$ method. The mean uncertainty in $\larg_n$ is given by the black square in the upper left corner.}
  \label{fig:largeur_met_sansTeff}
\end{figure}

Since $\larg_n$ strongly depend on the $\Teff$ values, the suppression of this dependence is necessary in order to assess the existence of a metallicity influence on the mode width. We therefore use the fits from Sec.\,\ref{Temperature_dependence} to correct the mode widths for the influence of $\Teff$ previously characterized in this section. The results, shown on Fig. \ref{fig:largeur_met_sansTeff} for the $\Kallinger$ method, exhibit no significant trend with the metallicity. Therefore, we conclude that we could not highlight a metallicity influence on the mode width with the precision we currently have on this parameter.

\listofobjects
\end{document}